\documentstyle[aps,epsf]{revtex}   

\def\bra{\langle}
\def\ket{\rangle}

\begin{document}

\title{Development of particle multiplicity distributions using a general form of 
the grand canonical partition function and applications to L3 and H1 Data}

\author{S.J. Lee}
\address{Department of Physics and Institute of Natural Sciences,
                    Kyung Hee University, Suwon, KyungGiDo, Korea}

\author{A.Z. Mekjian}
\address{Department of Physics, Rutgers University,
                    Piscataway, New Jersey}

\maketitle

\begin{abstract}

Various phenomenological models of particle multiplicity distributions are
discussed using a general form of a unified model which is based on the
grand canonical partition function and Feynman's path integral approach to
statistical processes. These models can be written as special cases of a
more general distribution which has three control parameters which are
$a$, $x$, $z$. The relation to these parameters to various physical quantities
are discussed. A connection of the parameter $a$ with Fisher's critical
exponent $\tau$ is developed. Using this grand canonical approach, moments,
cumulants and combinants are discussed and a physical interpretation of
the combinants are given and their behavior connected to the critical
exponent $\tau$. Various physical phenomena such as hierarchical structure,
void scaling relations, KNO scaling features, clan variables, and
branching laws are shown in terms of this general approach. Several
of these features which were previously developed in terms of the negative
binomial distribution are found to be more general. Both hierarchical
structure and void scaling relations depend on the Fisher exponent $\tau$.
Applications of our approach to the charged particle multiplicity distribution
in jets of L3 and H1 data are given.
It is shown that just looking at the mean and fluctuation of data is not
enough to distinguish these distributions or the underlying mechanism.
The mean, fluctuation and third cummulant of distribution determine three
parameters $x$, $z$, $a$.
We find that a generalized random work model fits the data better than
the widely used negative binomial model.

\end{abstract}



\section{Introduction}

Pion multiplicity distribution and their associated fluctuation and 
correlations have been of interest for several reasons.
Several models predict large fluctuations such as the disoriented
chiral condensate \cite{ref3,ref4}
and in density fluctuations from droplets arising in a first order
phase transition \cite{ref5}.
A well known procedure for studying correlations uses the
Bose-Einstein symmetries associated with pions in
a Hanbury Brown-Twiss analysis \cite{ref6}.
Such an analysis gives information about the space time
history of the collision through measurements of source parameters.
If the density of pions becomes large, Bose-Einstein correlation
may also lead to a strongly emitting system which has been
called a pion laser \cite{ref7}.
The pion laser model has been recently solved analytically
by T. Cs\"{o}rg\"{o} and J. Zimanyi \cite{ref7a}.
The importance of Bose-Einstein correlations has also been
illustrated in the observation of a condensation of atoms in
a harmonic oscillator or laser trap \cite{ref8}.
Previous interest in pionic distributions have centered around
the possibility of intermittency behavior \cite{ref9} and
fractal structure based on parallels with turbulent flow in fluids.
A distribution widely used to discuss such features has been
the negative binomial (NB) distribution \cite{ref10}
with its associated clan structure \cite{ref11,ref12}
and KNO scaling feature \cite{ref13}.
KNO scaling properties have been interpreted in terms of a
phase transition associated with a Feynman-Wilson gas \cite{ref14}.
Various other issues associated with pions include evidence for 
thermalization \cite{ref15}, critical point fluctuations \cite{ref16,ref17},
fluctuations from a first order phase transition \cite{ref18},
charge particle ratios and question of chemical equilibrium \cite{ref19},
the behavior of fluctuations in net charge in a QG plasma for
transition \cite{ref20,ref21}.

For lower energy heavy ion collisions, multifragmentation of nuclei
takes place. The fragment distribution can also be described
statistically by considering all the possible partition of $A$
nucleons into smaller clusters \cite{frag,canon}. 
This study gives a tool
for the description of nuclear multifragmentation distributions \cite{massd}, 
nuclear liquid-gas phase transition \cite{lgpha}, 
critical exponent, intermittency, and chaotic behavior \cite{scale,power}
of nuclear multifragmentation.
The same model can describe pionic distribution.
This possibility arises in our approach which is based on Feynman path 
integral methods where symmetrization of $A$ bosons or anti-symmetrization
of $A$ fermions leads to a cycle class decomposition of the
permutations associated with these symmetries.
The correspondance comes from the identification of clusters of size $k$
and the cycles of length $k$ in a permutation as discussed below.
In nuclear multifragmentation the number of nucleons, $A$, is fixed and
thus a canonical partition function approach is appropriate.
By contrast, the number of pions, $A$, in particle production
is not fixed and thus we need to use a grand canonical partition.
Using this parallel, we also note
that some results from cluster yields can be carried over into particle
multiplicity distributions and associated properties.  Namely,
we show the importance of the Fisher critical exponent $\tau$ and relate
it to one of the parameters called $a$ in our approach which has three
main parameters $a$, $x$, $z$. Moreover, taking special values of $a$, or
equivalently $\tau$, reduces our unified model to various specific
cases that are frequently used in particle production phenomenology.
Quantities that appear in this development can also be related
combinants as will be discussed.
In so doing we can give a physical
significance to the combinants and show how the Fisher exponent appears
in them and how the resulting hierarchical structure and void scaling
relations also depend on its value.

In next section a summary of a generalized statistical model
based on a grand canonical partition function will be given.
This generalized model can be used to study multiplicity 
distributions associated with particle production such as at RHIC.
Various models of particle multiplicity distribution using the 
grand canonical partition function will then be developed.
In Section II-B, we will summarize the multiplicity distribution and 
its various moments of all order for a general grand canonical 
ensemble which we will then use for particle production.
The physical meaning of the parameters used in this model is given here.
The differences and relations between the canonical nuclear multifragmentation 
and the grand canonical multiparticle production are sumarized in Sect. II-C.
In Sect. II-D and E, moments (especially the mean and the fluctuation)
in the general grand canonical model are related to variables used 
in other existing standard models in describing multiparticle production.
The connection of the general grand canonical model developed here
to existing standard models are summarized in Sect. II-F.
Moreover, we derive a generalized model (HGa) of the grand canonical  
partition which can further be reduced to a geometric, negative binomial,
and Lorentz/Catalan model.
Various moments and multiplicity distribution for a generalized HGa model
are summarized in Sect. II-G 
and another generalized model (GRW1D) in Sect. II-H. 
In Sect. \ref{sectcomp}, we compare various statistical properties between
different models within the generalized HGa model
and with data for charged particle distribution in jets of L3 and H1.
Fluctuation and multiplicity distribution in HGa are compared
with other standard models of multiparticle production.

\section{Generalized probability distributions}

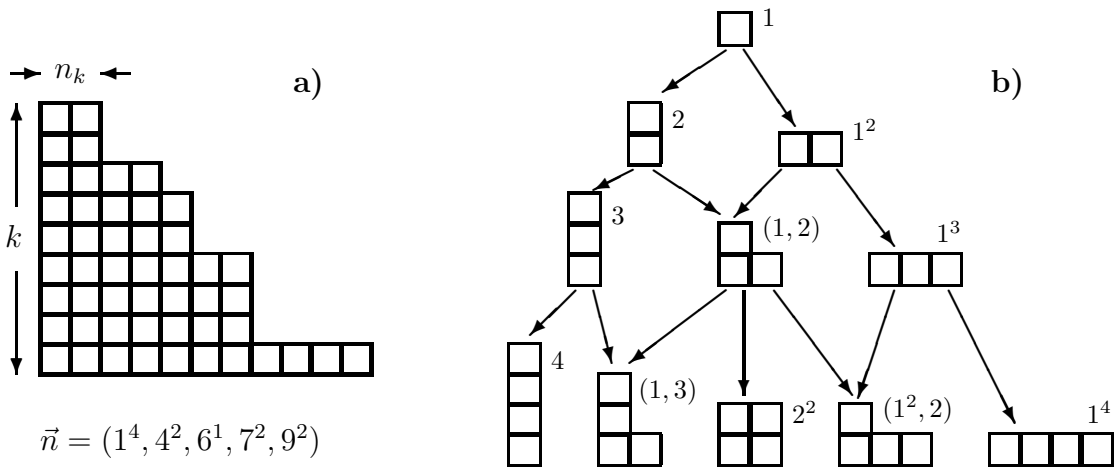
\begin{figure}[hbt]

\begin{center}

\setlength{\unitlength}{0.8cm}  
\begin{picture}(25,8)(7.5,0.0)     

\thicklines

\newsavebox{\diagbox}
\savebox{\diagbox}(0.5,0.5){ }

\put(12.2,6.2){\large\bf a)}

\put(8.0,0.25){\large\bf $\vec n = (1^4,4^2,6^1,7^2,9^2)$}

\put(8.01, 1.51){\line(0,1){4.5}}
\multiput(8.0, 5.5)(0.5,0){2}{\frame{\usebox{\diagbox}}}
\multiput(8.0, 5.0)(0.5,0){2}{\frame{\usebox{\diagbox}}}
\multiput(8.0, 4.5)(0.5,0){4}{\frame{\usebox{\diagbox}}}
\multiput(8.0, 4.0)(0.5,0){5}{\frame{\usebox{\diagbox}}}
\multiput(8.0, 3.5)(0.5,0){5}{\frame{\usebox{\diagbox}}}
\multiput(8.0, 3.0)(0.5,0){7}{\frame{\usebox{\diagbox}}}
\multiput(8.0, 2.5)(0.5,0){7}{\frame{\usebox{\diagbox}}}
\multiput(8.0, 2.0)(0.5,0){7}{\frame{\usebox{\diagbox}}}
\multiput(8.0, 1.5)(0.5,0){11}{\frame{\usebox{\diagbox}}}
\put(8.99, 5.0){\line(0,1){1.0}}
\put(9.99, 4.5){\line(0,1){0.5}}
\put(10.49, 3.5){\line(0,1){1.0}}
\put(11.49, 2.0){\line(0,1){1.5}}
\put(13.49, 1.5){\line(0,1){0.5}}
\put(8.01, 1.51){\line(1,0){5.5}}
\put(11.51, 2.01){\line(1,0){2.0}}
\put(10.51, 3.51){\line(1,0){1.0}}
\put(10.01, 4.51){\line(1,0){0.5}}
\put(9.01, 5.01){\line(1,0){1.0}}
\put(8.01, 6.01){\line(1,0){1.0}}
\multiput(7.5,6.5)(0,0.01){2}{\vector(1,0){0.5}}
\multiput(9.5,6.5)(0,0.01){2}{\vector(-1,0){0.5}}
\put(8.0,6.2){\makebox(1.0,0.6){\large\bf $n_k$}}
\multiput(7.6,3.2)(0,0.01){2}{\vector(0,-1){1.7}}
\multiput(7.6,4.3)(0,0.01){2}{\vector(0,1){1.7}}
\put(7.4,3.5){\makebox(0.3,0.6)[r]{\large\bf $k$}}
\put(15.5,0){
\setlength{\unitlength}{0.8cm}  
\begin{picture}(10,8)(0,0)
\thicklines

\savebox{\diagbox}(0.5,0.5){ }

\put(8,6.2){\large\bf b)}

\put(3.5, 7.0){\frame{\usebox{\diagbox}}}
\put(4.2,7.3){$1$}
\put(3.6,6.9){\vector(-3,-2){1.05}}
\put(3.9,6.9){\vector(2,-3){0.85}}

\multiput(2.0, 5)(0,0.5){2}{\frame{\usebox{\diagbox}}}
\put(2.7,5.6){$2$}
\put(2.05,4.9){\vector(-2,-1){0.65}}
\put(2.4,4.9){\vector(3,-2){1.1}}

\multiput(4.5, 5.0)(0.5,0){2}{\frame{\usebox{\diagbox}}}
\put(5.7,5.4){$1^2$}
\put(4.52,4.9){\vector(-1,-1){0.78}}
\put(5.45,4.9){\vector(3,-4){0.95}}

\multiput(1., 3.0)(0,0.5){3}{\frame{\usebox{\diagbox}}}
\put(1.7,4.0){$3$}
\put(1.1,2.9){\vector(-1,-1){0.75}}
\put(1.4,2.9){\vector(1,-4){0.3}}

\multiput(3.5, 3.0)(0,0.5){2}{\frame{\usebox{\diagbox}}}
\put(4.0, 3.0){\frame{\usebox{\diagbox}}}
\put(4.2,3.8){$(1,2)$}
\put(3.6,2.9){\vector(-4,-3){1.6}}
\put(3.9,2.9){\vector(0,-1){1.75}}
\put(4.4,2.9){\vector(3,-4){1.3}}

\multiput(6.0, 3.0)(0.5,0){3}{\frame{\usebox{\diagbox}}}
\put(7.1,3.7){$1^3$}
\put(6.4,2.9){\vector(-1,-3){0.6}}
\put(7.3,2.9){\vector(1,-2){1.1}}

\multiput(0., 0.0)(0,0.5){4}{\frame{\usebox{\diagbox}}}
\put(0.7,1.6){$4$}

\multiput(1.5, 0.0)(0,0.5){3}{\frame{\usebox{\diagbox}}}
\put(2.0, 0.0){\frame{\usebox{\diagbox}}}
\put(2.15,1.1){$(1,3)$}

\multiput(3.5, 0.0)(0,0.5){2}{\frame{\usebox{\diagbox}}}
\multiput(4.0, 0.0)(0,0.5){2}{\frame{\usebox{\diagbox}}}
\put(4.7,0.7){$2^2$}

\multiput(5.5, 0.0)(0.5,0){3}{\frame{\usebox{\diagbox}}}
\put(5.5, 0.5){\frame{\usebox{\diagbox}}}
\put(6.2,0.8){$(1^2,2)$}

\multiput(8.0, 0.0)(0.5,0){4}{\frame{\usebox{\diagbox}}}
\put(9.6,0.7){$1^4$}

\end{picture}
}

\end{picture}

\caption{Building partitions with blocks.
}
  \label{fig1}

\end{center}

\end{figure}

Consider a system composed of $N$ different types of species
or objects which could be the fragments 
in a fragmentation or in a cycle class description.
Any event of such a system can be associated with a vector
 $\vec n = \{n_k\} = (n_1, n_2, \cdots, n_k, \cdots, n_N)$
or $1^{n_1} 2^{n_2} 3^{n_3} \cdots k^{n_k} \cdots N^{n_N}$
where the non-negative integer $n_k$ is the number of individuals 
of species $k$.
For example $n_k$ can be the number of clusters of size $k$
or the number of cycles of length $k$ in a given permutation
of $n$ particles.
The later is important for Bose-Einstein and Fermi-Dirac statistics 
and particle multiplicity distributions.
A general block picture of $\vec n$ is shown in Fig.\ref{fig1}a.
Fig.\ref{fig1}b shows how the various partition can be developed
as an evolution from successively smaller systems.
The number of species $N$ can be infinity in general.

Various probability distributions related with this system can be 
developed by assigning an appropriate weight $x_k$ to each type $k$.
A weight $W(\vec x, \vec n)$ is then given to each event $\vec n$
and the type of weight that will be considered has the structure:
\begin{eqnarray}
 W(\vec x, \vec n) &=& \prod_{k=1}^N \left[\frac{x_k^{n_k}}{n_k!}\right]
      \label{weight}
\end{eqnarray}
The $n_k!$ are Gibbs factorials. 
Such a weight structure appears in Feynman's path integral approach and Bose-Einstein 
problems \cite{zarecur}. 
The $x_k$ will be given below and contains various physical quantities.
In Sect.\ref{sectIIb} we will show that the $x_k$'s are also the combinants 
which in turn can be related to the factorial cumulants.
Summing the weight $W(\vec x, \vec n)$ over all the possible
events of $\vec n$, the grand canonical partition function $Z(\vec x)$
of the system is given as
\begin{eqnarray}
 Z(\vec x) &=& \sum_{\vec n} W(\vec x, \vec n)
   = \sum_{\vec n} \prod_{k=1}^N \left[\frac{x_k^{n_k}}{n_k!}\right]
   = \exp\left[\sum_{k=1}^N x_k\right]
        \label{grandptf}  
\end{eqnarray}
The last equation holds due to the form of Eq.(\ref{weight})
of the weight $W(\vec x, \vec n)$, i.e., the factor $x_k^{n_k}/n_k!$ 
is the $n_k$-th order expansion term of $e^{x_k}$ .

Introducing other quantities $\alpha_k$ to each individual entity
or group of type $k$,
 $\vec\alpha = \{\alpha_k\}
     = (\alpha_1, \alpha_2, \cdots, \alpha_k, \cdots, \alpha_N)$,
we can define a canonical partition function $Z_A(\vec x)$ with a fixed $A$ as
\begin{eqnarray}
 A &=& \sum_{k=1}^N \alpha_k n_k = \vec\alpha\cdot\vec n
            \label{asum}  \\
 Z_A(\vec x) &=& \sum_{\vec n_A} W(\vec x, \vec n)
   = \sum_{\vec n_A} \prod_{k=1}^N \left[\frac{x_k^{n_k}}{n_k!}\right]
            \label{canoptf}   
\end{eqnarray}
with $Z(\vec x) = \sum_A Z_A(\vec x)$.
Here $\sum_{{\vec n}_A}$ is the summation over all events with
a fixed value of $A$, i.e., over a canonical ensemble of a fixed $A$,
and the $\sum_A$ is a summation over all the possible values of $A$;
it becomes $\sum_{A=0}^\infty$ for the case of $\alpha_k = k$ with
positive integer $k$.
The multiplicity of a partitions is
\begin{eqnarray}
 M &=& \sum_{k=1}^N n_k    \label{multip} 
\end{eqnarray}
The case $\alpha_k = k$ is encountered in fragmentation problems 
and permutation problems.
Various type of $\alpha_k$ can be used depending on the physics
related with the quantity $A$ as discussed in Ref.\cite{canon}.
The physics of the canonical ensemble depends on the choice of
the quantity $\alpha_k$ \cite{frag,canon}.
If we take $x_k \propto z^{\alpha_k}$, then the canonical partition
function $Z_A(\vec x)$ is the $z^A$ dependent term  
of the grand canonical partition function $Z(\vec x)$.
If we take $x_k \propto x$, then $x$ counts the number of clusters $M$ explicitly 
and the $x^M$ dependent term of the grand canonical partition function $Z(\vec x)$
is the partition function for events with multiplicity $M$.
There always is at least one event having $A = 0$, i.e., the
event where all $n_k$'s are zero, $\vec n = \vec 0$.
Thus $Z_0(\vec x) = 1$ if all $\alpha_k$ are non zero positive
since then there are no other possible events having $A = 0$.
Due to the form of the weight $W(\vec x, \vec n)$ given by Eq.(\ref{weight})
the canonical partition function $Z_A(\vec x)$  
satisfies a recurrence relation \cite{scale,zarecur}   
\begin{eqnarray}
 Z_A(\vec x) &=& \frac{1}{A} \sum_k \alpha_k x_k Z_{A-\alpha_k}(\vec x)
          \label{recur}
\end{eqnarray}
This relation is nothing but the constraint Eq.(\ref{asum})
in terms of the mean $\bra n_k\ket_A$ using Eq.(\ref{factma}).
For non-zero positive $\alpha_k$, there is no case having $A < 0$, i.e.,
$Z_A = 0$ for $A < 0$ 
and thus the $Z_A$ can be obtained by the recurence relation of
Eq.(\ref{recur}) starting from $Z_0(\vec x) = 1$.

\subsection{Probability distribution}

In a canonical ensemble of fixed $A$, we can define a probability
distribution of a specific partition $\vec n$ as
\begin{eqnarray}
 P_A(\vec x, \vec n) &=& \frac{W(\vec x, \vec n)}{Z_A(\vec x)}
\end{eqnarray}
With this probability, various mean values, fluctuations
and correlations of the number of species $n_k$ can be evaluated 
as a ratio of canonical partition functions for two different values 
of $A$ such as \cite{frag,canon}
\begin{eqnarray}
 \left<\frac{n_k!}{(n_k-m)!} \frac{n_j!}{(n_j-l)!}\right>_A
   &=& \sum_{\vec n_A} \frac{n_k!}{(n_k-m)!}
            \frac{n_j!}{(n_j-l)!} P_A(\vec x, \vec n)
    = x_k^m x_j^l \frac{Z_{A - m \alpha_k - l \alpha_j}(\vec x)}{Z_A(\vec x)}
          \label{factma}
\end{eqnarray}
Thus we have
 $\bra n_k\ket_A = x_k \frac{Z_{A-\alpha_k}(\vec x)}{Z_A(\vec x)}$.
The recurrence relation of Eq.(\ref{recur}) then follows simply
using the fact that $A = \sum_{k=1}^N \alpha_k \bra n_k\ket$. 
This distribution has been used in describing various fragment 
distributions in nuclear multifragmentation \cite{frag,canon,massd}.

Now knowing the partition functions $Z(\vec x)$ and $Z_A(\vec x)$,
we can associate a probability $P_A(\vec x)$ of the system to have a 
fixed value of $A$ in a grand canonical ensemble as
\begin{eqnarray}
 P_A(\vec x) &=& \frac{Z_A(\vec x)}{Z(\vec x)}
    = \frac{1}{Z(\vec x)} \frac{1}{\Gamma(A+1)}
      \left[\left(\frac{d}{d z}\right)^A Z(\vec x, z)\right]_{z=0} 
                 \label{pax}  \\
 Z(\vec x, z) &=& \sum_A Z_A(\vec x) z^A
    = \exp\left[\sum_k x_k z^{\alpha_k}\right]   \label{grandzxz}
\end{eqnarray}
The last step follows from the fact that 
the $z^A$ power term of $Z(\vec x, z)$ is $Z_A(\vec x)$
if we put $x_k = z^{\alpha_k}$.  
For particle production $P_A(\vec x)$ is the probability of having $A$ particles.
Thus the generating function $Z(\vec x, z)$ of $P_A$ can also
be looked at as a grand canonical partition function with
the weight $x_k$ replaced to be $x_k z^{\alpha_k}$,
where the variable $z$ counts $A$ explicitly
and $Z(\vec x, 1) = Z(\vec x)$.
If we consider $P_A(\vec x, z) = P_A(\vec x) z^A$ then $z$ has two roles;
one as a weight which is assigned the same to each constituent and 
another as a generating  parameter of the probability $P_A(\vec x)$.
For the case that $Z_0(\vec x) = 1$, the void probability $P_0$ \cite{hiera}
is the inverse of the grand canonical partition function, i.e.,
\begin{eqnarray}
 P_0(\vec x) &=& \frac{Z_0(\vec x)}{Z(\vec x)} = Z^{-1}(\vec x)
\end{eqnarray}
Inversely the canonical partition function $Z_A(\vec x)$ is the 
probability $P_A$ normalized or rescaled by the void probability $P_0$, i.e.,
\begin{eqnarray}
 Z_A(\vec x) &=& \frac{1}{\Gamma(A+1)} \left[\left(\frac{d}{d z}\right)^A
         Z(\vec x, z)\right]_{z=0}
   = \frac{P_A(\vec x)}{P_0(\vec x)}
\end{eqnarray}
Another type of generating function of $P_A$ that is frequently used
may be defined as
\begin{eqnarray}
 G(\vec x, u) &=& \sum_{A} P_A(\vec x) (1-u)^A
   = \frac{1}{Z(\vec x)} \sum_{A}^\infty Z_A(\vec x) (1-u)^A
         \nonumber  \\
  &=& \frac{Z(\vec x, z=1-u)}{Z(\vec x)}
   = \exp\left[\sum_k x_k [(1-u)^{\alpha_k} - 1] \right]
           \label{genu}   \\
 P_A(\vec x) &=& \frac{1}{\Gamma(A+1)} \left[\left(-\frac{d}{d u}\right)^A
        G(\vec x, u)\right]_{u=1}
\end{eqnarray}
We see that $G(\vec x, 0) = 1$,
$G(\vec x, 1) = P_0(\vec x) = Z_0(\vec x)/Z(\vec x)$
and $G(\vec x, 1-z) = Z(\vec x, z)/Z(\vec x)$.
Once the probability $P_A(\vec x)$ is determined, 
various statistical quantities can be evaluated.
Also the grand canonical partition function $Z(\vec x,z)$ is given
once we know the thermodynamic grand potential 
\begin{eqnarray}
 \Omega(\vec x, z) = - \ln Z(\vec x, z) = - \sum_k x_k z^{\alpha_k}.
               \label{grandpot}
\end{eqnarray}
This result can be used to study the statistical properties of a system.
Moreover various moments and cumulants, mean values and fluctuations may
be obtained using the generating function \cite{frag,canon}.

\subsection{Moments and cumulants; 
combinants and hierachical structure in grand canonical ensemble}
   \label{sectIIb} 

This subsection gives general expression for various quantities
that will be used later when we discuss specific models.
Since the probability of a specific event $\vec n$ in a grand canonical
ensemble is given by
\begin{eqnarray}
 P(\vec x, \vec n) &=& \frac{W(\vec x, \vec n)}{Z(\vec x)}
    = \sum_A P_A(\vec x) P_A(\vec x, \vec n)
\end{eqnarray}
the mean of a quantity $F$ in a grand canonical ensemble is related 
to the mean of $F$ in a canonical ensemble as
\begin{eqnarray}
 \bra F\ket &=& \sum_{\vec n} F P(\vec x, \vec n)
  = \sum_A P_A(\vec x) \sum_{\vec n_A} F P_A(\vec x, \vec n)
   = \sum_A P_A(\vec x) \bra F\ket_A
\end{eqnarray}
We can easily show that
\begin{eqnarray} 
 \bra n_k\ket &=& \sum_{\vec n} n_k P(\vec x, \vec n)
   =\sum_A P_A(\vec x) \bra n_k\ket_A
   = x_k   \label{meank}     \\
 \bra M\ket &=& \sum_{\vec n} \left(\sum_k n_k\right) P(\vec x, \vec n)
   = \sum_k x_k
   = - \Omega(\vec x, z=1)
\end{eqnarray}
This result shows that the weight factor $x_k$ in this model
is the mean number $\bra n_k\ket$ in a grand canonical ensemble and
that the mean multiplicity $\bra M\ket$ is related to the thermodynamic potential.
The $m$-th power moment of $A$
and its factorial moments are given simply by
\begin{eqnarray}
 \bra A^m\ket(\vec x) &\equiv& \sum_{\vec n}
           \left(\sum_{k=1}^N \alpha_k n_k\right)^m P(\vec x, \vec n)
  =  \frac{1}{Z(\vec x)} \left[\left(z \frac{d}{d z}\right)^m
          Z(\vec x, z)\right]_{z=1}       \\
 \left<\frac{\Gamma(A+1)}{\Gamma(A-m+1)}\right>(\vec x)
  &\equiv& \sum_{\vec n} \frac{\Gamma(A+1)}{\Gamma(A-m+1)} P(\vec x, \vec n)
   = \frac{1}{Z(\vec x)} \left[\left(\frac{d}{d z}\right)^m
          Z(\vec x, z)\right]_{z=1}      
\end{eqnarray}
Similarly the $m$-th cumulants , which is the power moments of $\alpha_k$,
and the factoral cumulants are
\begin{eqnarray}
 \bra\alpha^m\ket(\vec x) &\equiv& \left<\sum_{k=1}^N \alpha_k^m n_k\right>
   = \left[\left(z \frac{d}{d z}\right)^m \ln Z(\vec x, z)\right]_{z=1}
   = \sum_{k=1}^\infty \alpha_k^m x_k     \label{alpham}  \\
 f_m(\vec x) &\equiv& \left< \sum_{k=1}^N
          \frac{\Gamma(\alpha_k+1)}{\Gamma(\alpha_k-m+1)} n_k\right>
   = \left[\left(\frac{d}{d z}\right)^m \ln Z(\vec x, z)\right]_{z=1}
              \nonumber   \\
  &=& \left[\left(- \frac{d}{d u}\right)^m \ln G(\vec x, u)\right]_{u=0}
   = \sum_{k=1}^N \frac{\Gamma(\alpha_k+1)}{\Gamma(\alpha_k-m+1)} x_k 
               \label{factfm} 
\end{eqnarray}
Due to Eqs.(\ref{grandzxz}) and (\ref{grandpot})  
we can see easily that, for the power moments of $\alpha_k$,
\begin{eqnarray}
 \bra\alpha^0\ket &=& \sum_{k=1}^N x_k = f_0 = \bra M\ket    \\
 \bra\alpha^1\ket &=& \sum_{k=1}^N \alpha_k x_k = f_1 = \bra A\ket   \label{malph1} \\
 \bra\alpha^2\ket &=& \sum_{k=1}^N \alpha_k^2 x_k
    = f_2 + f_1 = \bra(A - \bra A\ket)^2\ket = \bra A^2\ket - \bra A\ket^2 = \sigma^2
                        \label{malph2}     \\
 \bra\alpha^3\ket &=& \sum_{k=1}^N \alpha_k^3 x_k
           = f_3 + 3 f_2 + f_1 = \bra(A - \bra A\ket)^3\ket     \label{malph3}    
\end{eqnarray}
The power moments of $\alpha_k$ are directly related to the
power moments of $A$ measured from the mean $\bra A\ket$, i.e.,
the cumulants $\bra\alpha^m\ket$ are same with the central moments of $A$.
This simple relation does not hold for $m \ge 4$
but we can evaluate them starting from $\bra\alpha^3\ket$ 
using the recurrence relation 
\begin{eqnarray}
 \bra\alpha^{m+1}\ket(\vec x, z) &=& \left(z \frac{d}{d z}\right)^{m+1} \ln Z(\vec x, z)
   = \left(z \frac{d}{d z}\right) \bra\alpha^m\ket(\vec x, z)
\end{eqnarray}
Similarly the $m$-th factorial cumulants $f_m$, which is the factorial moments
of $\alpha_k$, can be found using recurrence relation
\begin{eqnarray}
 \frac{f_{m+1}(\vec x, z)}{f_m(\vec x, z)}
    &=& z \left(\frac{d}{d z}\right) \ln f_m(\vec x, z)  -  m
               \ \ \ {\rm or} \ \ \
 \frac{f_m(\vec x, z)}{z^m}
    = \frac{d}{d z} \left(\frac{f_{m-1}(\vec x, z)}{z^{m-1}}\right)
    = \left(\frac{d}{d z}\right)^m f_0(\vec x, z)        \label{recfm}
\end{eqnarray}
starting from 
\begin{eqnarray}
 f_0(\vec x, z) &=& \bra M\ket(\vec x, z) = \ln Z(\vec x, z) = - \Omega(\vec x, z)
    = \sum_{k=1}^N x_k z^{\alpha_k}    \\
 f_1(\vec x, z) &=& \bra A\ket(\vec x, z) = \sum_{k=1}^N \alpha_k x_k z^{\alpha_k}
\end{eqnarray}
The reduced factorial cumulants $\kappa_m$ defined in Ref.\cite{hiera}
corresponds to the factorial cumulants $f_m$ normalized with mean
number $\bra A\ket = \bar A$ as
\begin{eqnarray}
 \kappa_m(\vec x, z) &=& \frac{f_m(\vec x, z)}{{\bar A}^m}
\end{eqnarray}
Thus with $\kappa_1 \equiv 1$ and $\kappa_0 = f_0 = \bra M\ket$.
The cummulants are directly related with the generating functions as
\begin{eqnarray}
 G(\vec x, u) &=& \exp\left[\sum_{m=1}^\infty \frac{(-u)^m}{m!} f_m(\vec x)\right]
   = \exp\left[\sum_{m=1}^\infty \frac{(\ln[1-u])^m}{m!} \bra\alpha^m\ket(\vec x)\right]      
              \\  
 Z(\vec x, z) &=& \exp\left[\sum_{m=0}^\infty \frac{(z-1)^m}{m!} f_m(\vec x)\right]
    = \exp\left[\sum_{m=0}^\infty \frac{(\ln z)^m}{m!} \bra\alpha^m\ket(\vec x)\right]
\end{eqnarray}
The generating function $Z(\vec x, z)$ differs from the generating
function $G(\vec x, u=1-z)$ only by an extra term of $m = 0$ in their exponent.

The relation between the $x_k$'s and the $Z$ shows that
the $x_k$'s are also the combinants of Ref.\cite{refgyu}.
In the approach presented here,
the combinants are given an underlying significance through the partition
weight $W(\vec x, \vec n)$ of Eq.(\ref{weight}).
In turn the combinants $x_k$ can be related to the factorial cumulants $f_m$
defined by
\begin{eqnarray}
 \ln Z(\vec x, z) &=& \sum_k x_k z^{\alpha_k}
   = \sum_{m=0}^\infty \frac{(z-1)^m}{m!} f_m(\vec x)
\end{eqnarray}
The factorial cumulants $f_m$ are the $m$-th order factorial moments 
of $\alpha_k$ of Eq.(\ref{factfm}). Thus  
\begin{eqnarray}
 f_m &=& m! \sum_{k=m}^\infty \pmatrix{k \cr m} x_k
\end{eqnarray}
for $\alpha_k = k$ which is the Eq.(\ref{factfm}) with $N \to \infty$.
The normalized factorial cumulant, i.e., the reduced cumulant, is
\begin{eqnarray}
 \kappa_m &=& f_m/ \bar A^m
    = (m-1)! \kappa_2^{m-1}    \label{kappam}
\end{eqnarray}
for a negative binomial (NB) distribution,
This result of Eq.(\ref{kappam}) shows  that $\kappa_m$ for NB has 
an hierarchical structure of a distribution at the reduced cumulant level
which was realized for the NB distribution in Ref.\cite{hiera}.
This result will be generalized later.

Also using the above power moments and factorial moments we can study
voids and void scaling relation, hierarchical structure, 
combinant and cummulant properties
which will be discussed below.

\subsection{Multi-fragmentation versus multiparticle production
and the Fisher exponent for particle production }  
   \label{sectfrag}

Since our approach was first used to discuss multifragmentation
and then later extended to include multiparticle production,
we briefly mention some of difference between multifragmentation
and multiparticle production.
We also show how the Fisher exponent $\tau$, which initially appeared in
cluster yields around a crititical point, may manifest itself in particle
production yields.

In nuclear multifragmentation, $\alpha_k = k$ is the number of nucleons
in a fragment and $n_k$ is the number of fragments of size $k$.
The total number of fragments is $M = \sum_k n_k$
and the total number of nucleons is $A = \sum_k k n_k$.
In nuclear multifragmentation, the total number of nucleons $A$
is usually fixed and we study the distribution in size $k$ 
of the mean multiplicity $\bra n_k\ket_A = x_k Z_{A-k}(\vec x)/Z_A(\vec x)$ of 
fragments of size $k$ in a canonical ensemble. 
In multiparticle production the $A$ is the total number (multiplicity) of 
produced particles and is not fixed and we study the multiplicity distribution 
of produced particles $P_A = Z_A(\vec x)/Z(\vec x)$ as a function of $A$ in a 
grand canonical ensemble.
The mean multiplicity $\bra M\ket_A$ and  $\bra n_k\ket_A$ 
in a canonical ensemble are related with grand canonical ensemble as
\begin{eqnarray}
 \bra n_k\ket &=& \sum_{A=0}^\infty \bra n_k\ket_A P_A = x_k     \nonumber \\
 \bra M\ket &=& \sum_{A=0}^\infty \bra M\ket_A P_A = \sum_k x_k  \nonumber \\
 \bra A\ket &=& \sum_{A=0}^\infty \sum_k k \bra n_k\ket_A P_A
      = \sum_{A=0}^\infty A P_A = \sum_k k x_k     \nonumber
\end{eqnarray}
The weight $x_k$ can be determined experimentally from the mean 
multiplicity $\bra n_k\ket$ of cluster size $k$ in a grand canonical ensemble.

In our initial fragmentation studies \cite{frag,canon,massd} 
we have used $x_k = xz^k/k$.
This choice gives $Z_A(x,z) = (z^A/A!) \Gamma(x+A)/\Gamma(x)$ and thus
$Z_{A-k}/Z_A \to z^{-k}$ as $A \to \infty$ where the $z = e^{\beta\mu}$ 
with $\mu$ the chemical potential. 
From Eq.(\ref{factma}), $\bra n_k\ket_A = x_k Z_{A-k}/Z_A$, thus $\bra n_k\ket_A \to x/k$ 
as $A \to \infty$ in the case of nuclear fragmentation in the canonical
ensemble of fixed $A$ for $x_k = xz^k/k$ and the 
yields fall as a power law $1/k$ at a critical point ($x=1$)
for this choice of $x_k$. 
The yield in the grand canonical ensemble is $\bra n_k\ket = x_k = x z^k/k$ and
this becomes the same as $\bra n_k\ket_A \to x/k$ in $A \to \infty$ limit 
for $z=1$ with $\mu=0$.
In general, cluster yields fall for large $k$ as a power law
$1/k^\tau$ at a critical point where $\tau$ is the Fisher critical exponent.
The above choice of $x_k$ can be generalized to $x_k = x z^k/k^\tau$ so that the
cluster yields fall as $1/k^\tau$.
The $\tau$ determines the grand canonical partition function
 $Z(\vec x) = \exp[\sum_k x_k] = \exp[x \sum_k z^k /k^\tau]$.
Forms for $x_k$ whose asymptotic behavior
is $x z^k/k^\tau$ can also be used. 
In section II.F we will give various
choice for $x_k$ for particle production. For example, the above choice
of $x_k$ with $\tau = 1$ gives the frequently used negative binomial distribution
with $x$ the negative binomial parameter which determines the degree of
depature from Poisson statistics. Other distributions can be generated
and their associated $\tau$ dependence will also be given. Then we will show
how the $\tau$ dependence of the combinants shows up in hierarchical
structure and void scaling relations in sect II.G and in other places.
The physical meaning of $x$ and $z$ will also be discussed for these other cases.
Since an exact description of particle multiplicity yields is not known,
we consider $\tau$ as a free exponent to be determined by comparing various
resulting distributions with experiment. This procedure is what is done
in cluster yields.
In multi-fragmentation distribution, the mean number of clusters $\bra M\ket$ is 
determined by $x$ and $z$ is directly related with the number of constituent
particles $A$. The large $k$ behavior of the mean number of cluster $\bra n_k\ket$
gives the Fisher power $\tau$.
In multiparticle distribution, $x$ and $z$ determine the mean particle number $\bra A\ket$
or the peak position of the particle number distribution $P_A$ and the fluctuation
 $\sigma^2 = \bra A^2\ket - \bra A\ket^2$ or the width of the particle distribution.
The Fisher power $\tau$ determines the large $k$ behavior of the weight $x_k$ and
offers a new parameter in particle production.
$\bra n_k\ket = x_k$ in grand canonical ensemble.

Bose-Einstein condensation of atoms in a box of sides $L$
of dimensions $d$ have $x = L^d/\lambda_T^d$
with $\lambda_T = h/(2\pi m k_B T)^{1/2}$
and have $\tau = 1 + d/2$.
Feynman used random walk arguments, the closing of a cycle parallels
a closed random walk, to discuss his choice of $x_k$ in his discussion
of a superfluid phase transition in liquid helium.
Thermal emission of pions based on statistical mechanics
and equilibrium ideas have been popular descriptions of pions
coming from relativistic heavy ion collisions.
For thermal models \cite{refmek}, the
 $x_k = (V T^3/2\pi^2) (m/T)^2 K_2(km/T)/k$
for a cycle length $k$ or a cluster of size $k$
with $K_2$ a Mac Donald function.
For low temperatures, 
 $x_k = (V/\lambda_T^3) (e^{-m/T})^k /k^{5/2}$ and the 
Boltzmann factor in mass, $e^{-km/T}$, suppresses large fluctuations.
In the high temperature limit and/or zero pion mass limit
 $x_k = (V/\pi^2) T^3 /k^4$. Thus the high temperature limit has $\tau = 4$ 
and the low temperature limit has $\tau=5/2$.
The $x_k$ can be used to generate the pion probability distribution $P_n$.
The thermal models can be combined with hydrodynamic descriptions
and an application was given \cite{refmek} to $158 A$ GeV Pb+Pb data measured
by the CERN/NA44 and CERN/NA49 collaborations.
The results of Ref.\cite{refmek} showed a Gaussian distribution
with a width about 20 \% larger than the Poisson result.

\subsection{Clan parameters and void parameters and void scaling relations}

Van Hove and Giovannini have introduced clan variables $N_c$ and $n_c$ 
to describe a general class of probability distributions,
with most discussions of these variables centering around
the negative binomial distribution \cite{ref12}.  
These variables are defined as
\begin{eqnarray}
 N_c &=& \bra M\ket = \ln Z = f_0,  \hspace{1cm}  n_c = \bra A\ket/N_c = f_1/f_0
\end{eqnarray}
where the mean number of clans is $N_c$ and the $n_c$ is the mean number of
members per clan.
The $Z$ is the grand canonical generating function 
and thus $N_c = \sum_k x_k$ 
where $x_k$ is the cycle class weight distribution $\vec x$.
The $\bra A\ket = \sum_k k \bra n_k\ket$ is the mean number of total members (particles).
The $n_k$ here is the number of clans of size $k$ having $k$ members.

The clan variable $N_c$ is also related to the void probability
 $P_0 = Z_0/Z = 1/Z = e^{-N_c}$; thus $N_c = - \ln P_0 = f_0$.
An important function in void analysis is
 $\chi = -\ln P_0/\bra A\ket = N_c/\bra A\ket = 1/n_c$.
Thus the void parameters, 
void probability $P_0$ and void function $\nu = \chi$ \cite{hiera},  
are equivalent to
the generalized clan parameters $N_c$ and $n_c$ 
with the equivalence given by:
\begin{eqnarray}
 f_0(\vec x) &=& \ln Z(\vec x) = -\ln P_0(\vec x) = N_c  \\
 \chi(\vec x) &\equiv& \frac{f_0(\vec x)}{\bra A\ket} 
    = \nu = {n_c}^{-1}
\end{eqnarray}
The $- \chi$ is the normalized grand potential $\Omega = -f_0$ 
for the mean $\bra A\ket$.

Void analysis looks for scaling properties associated with $\chi$;
specifically, $\chi$ is a function of the combination $\bra A\ket\xi$
where $\xi$ is the coefficient of $\bra A\ket^2$ in the fluctuation
 $\sigma^2 = \bra A\ket + \xi \bra A\ket^2$.
Since $f_2 = \bra\alpha^2\ket - \bra\alpha\ket = \bra(A - \bra A\ket)^2\ket - \bra A\ket$,
the variance of $A$ in a grand canonical ensemble becomes
\begin{eqnarray}
 \sigma^2 &\equiv& \bra A^2\ket - \bra A\ket^2 =\bra\alpha^2\ket = \sum_k \alpha_k^2 x_k
         \nonumber  \\
  &=& \bra A\ket + f_2 = \bra A\ket + \xi \bra A\ket^2
   = \bra A\ket \left[ 1 + \xi(\vec x) \bra A\ket\right]
\end{eqnarray}
with $\xi = \kappa_2$, i.e., the normalized factorial cumulant.  
Since the variance for a Poissonian distribution is the
same as the mean, $\sigma^2 = \bra A\ket$, the $\xi = 0$;
thus, the parameter $\xi \bra A\ket$ represents a degree of departure from
Poissonian fluctuation normalized by mean $\bra A\ket$
of the distribution.
A well known non-Poissonian example is a NB distribution
which has $\xi = \frac{1}{x}$ and 
this becomes Plank distribution with $x = 1$.
Thus the variable $x$ in $x_k$ determines the strength of the non-Poissonian
fluctuation term for the negative binomial distribution. In general,
$x$ and $z$ determine the mean particle number or the peak position of the
particle multiplicity distribution and the fluctuation or the width of the
particle distribution.
Using the recurrence relation Eq.(\ref{recfm}) for $f_m$, we can show that
\begin{eqnarray}
 \xi(\vec x,z) \bra A\ket(\vec x) &=& \chi^{-1}(\vec x, z)
      - z \left(\frac{d}{d z}\right) \ln \chi(\vec x, z) - 1
   = \frac{1 - z \left(\frac{d}{d z}\right) \chi(\vec x, z)}
          {\chi(\vec x, z)}  - 1     \\
 \kappa_3(\vec x, z) &=& \frac{f_3(\vec x, z)}{\bra A\ket^3}
    = \frac{\xi(\vec x, z)}{\bra A\ket} \left(z \frac{d}{d z}\right)
              \ln\left[\xi(\vec x, z) \bra A\ket^2\right]
        - 2 \frac{\xi(\vec x, z)}{\bra A\ket}  
\end{eqnarray}
A NB distribution has
 $\chi = \ln(1 + \xi\bra A\ket)/(\xi\bra A\ket)$
while the Lorentz/Catalan (LC) distribution 
discussed in Ref. \cite{prl86} and below
has $\chi = (\sqrt{2\xi\bra A\ket+1} - 1)/(\xi\bra A\ket)$.
We will study in Sect. \ref{voidscal}, 
$\chi$ {\it vs} $\xi\bra A\ket$, i.e., the void or clan variable {\it vs}
the fluctuation for various choices of $x_k$
summarized in Table \ref{tabl1}.

\subsection{Ancestral or evolutionary variables}  \label{ancest}

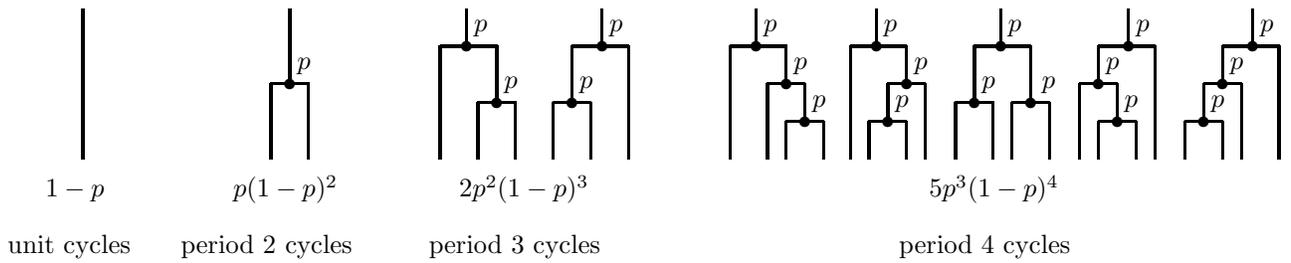
\begin{figure}[hbt]
\setlength{\unitlength}{0.5cm}  
\begin{center}
\begin{picture}(35,7)(1.2,4.0)  
  \thicklines
 \put(3,7){\line(0,1){4}}
    \put(2,6.0){$1-p$}  \put(1,4.5){unit cycles}
 \put(8.5,9){\line(0,1){2}}
    \put(8.0,9){\line(1,0){1}}  \put(8.5,9){\circle*{0.3}}
    \put(8.7,9.4){$p$}
    \put(8.0,7){\line(0,1){2}}  \put(9.0,7){\line(0,1){2}}
    \put(7.0,6.0){$p(1-p)^2$}   \put(5.6,4.5){period 2 cycles}
 \put(13.2,10){\line(0,1){1}}
    \put(12.5,10){\line(1,0){1.5}}  \put(13.2,10){\circle*{0.3}}
    \put(13.4,10.4){$p$}
    \put(12.5,7){\line(0,1){3}}  \put(14.0,8.5){\line(0,1){1.5}}
    \put(13.5,8.5){\line(1,0){1}}  \put(14.0,8.5){\circle*{0.3}}
    \put(14.2,8.9){$p$}
    \put(13.5,7){\line(0,1){1.5}}  \put(14.5,7.0){\line(0,1){1.5}}
   \put(16.8,10){\line(0,1){1}}
    \put(16.0,10){\line(1,0){1.5}}  \put(16.8,10){\circle*{0.3}}
    \put(17.0,10.4){$p$}
    \put(17.5,7){\line(0,1){3}}  \put(16.0,8.5){\line(0,1){1.5}}
    \put(15.5,8.5){\line(1,0){1}}  \put(16.0,8.5){\circle*{0.3}}
    \put(16.2,8.9){$p$}
    \put(15.5,7){\line(0,1){1.5}}  \put(16.5,7.0){\line(0,1){1.5}}
    \put(13.0,6.0){$2p^2(1-p)^3$}   \put(12.2,4.5){period 3 cycles}
 \put(20.9,10){\line(0,1){1}}
    \put(20.2,10){\line(1,0){1.5}}  \put(20.9,10){\circle*{0.3}}
    \put(21.1,10.4){$p$}
    \put(20.2,7){\line(0,1){3}}  \put(21.7,9.0){\line(0,1){1.0}}
    \put(21.2,9.0){\line(1,0){1}}  \put(21.7,9.0){\circle*{0.3}}
    \put(21.9,9.4){$p$}
    \put(21.2,7.0){\line(0,1){2.0}}  \put(22.2,8.0){\line(0,1){1.0}}
    \put(21.7,8.0){\line(1,0){1}}  \put(22.2,8.0){\circle*{0.3}}
    \put(22.4,8.4){$p$}
    \put(21.7,7.0){\line(0,1){1.0}}  \put(22.7,7.0){\line(0,1){1.0}}
  \put(24.1,10){\line(0,1){1}}
    \put(23.4,10){\line(1,0){1.5}}  \put(24.1,10){\circle*{0.3}}
    \put(24.3,10.4){$p$}
    \put(23.4,7){\line(0,1){3}}  \put(24.9,9.0){\line(0,1){1.0}}
    \put(24.4,9.0){\line(1,0){1}}  \put(24.9,9.0){\circle*{0.3}}
    \put(25.1,9.4){$p$}
    \put(24.4,8.0){\line(0,1){1.0}}  \put(25.4,7.0){\line(0,1){2.0}}
    \put(23.9,8.0){\line(1,0){1}}  \put(24.4,8.0){\circle*{0.3}}
    \put(24.6,8.4){$p$}
    \put(23.9,7.0){\line(0,1){1.0}}  \put(24.9,7.0){\line(0,1){1.0}}
  \put(27.4,10){\line(0,1){1}}
    \put(26.7,10){\line(1,0){1.5}}  \put(27.4,10){\circle*{0.3}}
    \put(27.6,10.4){$p$}
    \put(26.7,8.5){\line(0,1){1.5}}  \put(28.2,8.5){\line(0,1){1.5}}
    \put(26.2,8.5){\line(1,0){1}}  \put(26.7,8.5){\circle*{0.3}}
    \put(26.9,8.9){$p$}
    \put(26.2,7){\line(0,1){1.5}}  \put(27.2,7.0){\line(0,1){1.5}}
    \put(27.7,8.5){\line(1,0){1}}  \put(28.2,8.5){\circle*{0.3}}
    \put(28.5,8.9){$p$}
    \put(27.7,7){\line(0,1){1.5}}  \put(28.7,7.0){\line(0,1){1.5}}
   \put(30.8,10){\line(0,1){1}}
    \put(30.0,10){\line(1,0){1.5}}  \put(30.8,10){\circle*{0.3}}
    \put(31.0,10.4){$p$}
    \put(31.5,7){\line(0,1){3}}  \put(30.0,9.0){\line(0,1){1.0}}
    \put(29.5,9.0){\line(1,0){1}}  \put(30.0,9.0){\circle*{0.3}}
    \put(30.2,9.4){$p$}
    \put(29.5,7.0){\line(0,1){2.0}}  \put(30.5,8.0){\line(0,1){1.0}}
    \put(30.0,8.0){\line(1,0){1}}  \put(30.5,8.0){\circle*{0.3}}
    \put(30.7,8.4){$p$}
    \put(30.0,7.0){\line(0,1){1.0}}  \put(31.0,7.0){\line(0,1){1.0}}
   \put(34.1,10){\line(0,1){1}}
    \put(33.3,10){\line(1,0){1.5}}  \put(34.1,10){\circle*{0.3}}
    \put(34.3,10.4){$p$}
    \put(34.8,7){\line(0,1){3}}  \put(33.3,9.0){\line(0,1){1.0}}
    \put(32.8,9.0){\line(1,0){1}}  \put(33.3,9.0){\circle*{0.3}}
    \put(33.5,9.4){$p$}
    \put(32.8,8.0){\line(0,1){1.0}}  \put(33.8,7.0){\line(0,1){2.0}}
    \put(32.3,8.0){\line(1,0){1}}  \put(32.8,8.0){\circle*{0.3}}
    \put(33.0,8.4){$p$}
    \put(32.3,7.0){\line(0,1){1.0}}  \put(33.3,7.0){\line(0,1){1.0}}
   \put(25.5,6.0){$5p^3(1-p)^4$}   \put(24.7,4.5){period 4 cycles}
\end{picture}
\end{center}
\caption{Evolutionary lines of descent in a hierarchical topology.
Each branch increases the cycle length with probability $p$, survival $1-p$.
The probability distribution evolves from Poisson to chaotic.
For clusters each branch generates a bigger cluster.
  }

 \label{LCfig}
\end{figure}

The LC model, which has $x_k = \frac{1}{k} 2^{-2(k-1)} \pmatrix{2(k-1) \cr k-1} x z^k$,
was shown to be a useful model for discussing
an underlying splitting dynamics when ancestral or evolutionary
variables $p$ and $\beta$ were introduced into $x$ and $z$ as
discussed in Ref.\cite{prl86}.
Specifically $x = \beta/4p$ and $z = 4p(1-p)$
giving $x_k = \beta C_k p^{(k-1)} (1-p)^k$  
where $C_k = \frac{1}{k} \pmatrix{2(k-1) \cr k-1}$.
Percolation or splitting dynamics with
a branching probability $p$ and survival probability $(1-p)$
has a hierarchical topology as shown in Fig.\ref{LCfig}.
Weighting each diagram by $x_k = \beta C_k p^{k-1}(1-p)^k$,
the evolutionary or ancestral variables are related to 
the clan variables $N_c = \bra M\ket$ and $n_c = \bra A\ket/N_c$.
$C_k$ is the number of diagrams of size $k$ 
shown in Fig.\ref{LCfig}.
For this case the evolutionary dynamics is just that of the LC model.
Then with $\beta$ set equal to 1, $x_1 = (1-p)$,
$x_2 = p (1-p)^2$, $x_3 = 2 p^2 (1-p)^3$, $x_4 = 5 p^3 (1-p)^4$, etc.
The interpretation of this set of $x_k$'s  
reads as follows:
$x_1$ has 1 surviving line without a branch ($p^0(1-p)^1$) and 
one diagram ($C_1 = 1$),
$x_2$ has 1 branch ($p^1$) leading to 2 surviving lines ($(1-p)^2$)
and one diagram ($C_2 = 1$),
$x_3$ has 2 branch points ($p^2$) leading 3 surviving lines ($(1-p)^3$)
and two diagrams ($C_3 = 2$),
$x_4$ has 3 branch points($p^3$), 4 surviving lines ($(1-p)^4$) 
and 5 diagrams ($C_4 = 5$), etc.
In these evolutionary/ancestral variables 
the $f_0 = \sum x_k = 2 x (1 - \sqrt{1-z})$ which determines $Z$ is 
$f_0 = \beta$ for all $p \le 1/2$.
For $p \ge 1/2$, $f_0 = \sum x_k$ is no longer a constant
and is $f_0 = \beta (1-p)/p$.
To keep $f_0 = \sum_k x_k$ a constant 
without changing $f_1 = \bra n\ket = \sum_k k \bra n_k\ket = \beta \frac{(1-p)}{|1-2p|}$,
a $x_\infty = \phi_\infty$ was introduced in Ref.\cite{prl86}.
The $\phi_\infty = 0$ for $p \le 1/2$ 
and $\phi_\infty = \beta (2p-1)/p$ for $p \ge 1/2$.
For $p \ge 1/2$ there is a finite probability ($\phi_\infty \ne 0$) that 
the splitting will go on forever ($k \to \infty$).   
In percolation above a certain $p$ an infinite cluster is formed and 
$\phi_\infty$ is similar to the strength of the infinite cluster.
Moreover the sudden appearance of $\phi_\infty$ is similar
to the behavior of an order parameter in a phase transition.
The appearance of $\phi_\infty$ can also be interpreted as
the sudden appearance of a jet without pion ($k = 0$), 
if we introduce $x_0 = \phi_\infty$ instead of $x_\infty = \phi_\infty$.

Since the clan variables are $N_c = \bra M\ket = f_0$ and $n_c = \bra A\ket/N_c = f_1/f_0$,
then, for the LC model with evolutionary or ancestral varables $\beta$ and $p$,
\begin{eqnarray}
 N_c &=& \bra M\ket = f_0 = \beta \frac{1 - |1-2p|}{2p}         \nonumber \\
 n_c &=& \bra A\ket/N_c = \frac{2p(1-p)}{|1-2p|(1 - |1-2p|)}    \nonumber \\
 \bra A\ket &=& \beta \frac{(1-p)}{|1-2p|}                      \nonumber \\
 p &=& \frac{1}{2} \left[1 \mp \frac{1}{2 n_c - 1}\right]       \nonumber \\
 \beta &=& N_c \frac{n_c - (1 \pm 1)/2}{n_c - 1}                \nonumber
\end{eqnarray}
These can be reduced to $N_c = \beta$, $n_c = (1-p)/(1-2p)$,
and $2p = 1 - 1/(2n_c - 1)$ for $p \le 1/2$
while $N_c = \beta (1-p)/p$, $n_c = p/(2p-1)$, 
and $2p = 1 + 1/(2n_c - 1)$ for $p > 1/2$.
Since the branching probability $p$ varies in $0 \le p \le 1$,
the clan variable $n_c$ has $n_c \ge 1$
with $n_c = 1$ at $p = 0$ and 1, only one member per clan in average, 
and $n_c = \infty$ at $p = 1/2$, infinitly many members per clan.
The LC model thus connects the clan variable $n_c$ to the probability $p$ 
of branching in the evolutionary or ancestral picture of Fig.\ref{LCfig} 
or in a percolation model.
For Poisson processes $p=0$ (no branching ), 
$x_k = \beta \delta_{k1}$ (only unit cycles and no BE correlations) and $n_c = 1$
(one member in each clan in average).

A connection of the LC model can also be made with a Ginzburg-Landau
theory of phase transitions and a Feynman-Wilson Gas.
These connections are discussed in Ref. \cite{prc65}.
where $x$ and $z$ are related to coefficients in the Ginzburg-Landau approach.

\subsection{Unifying various distributions with Gauss hypergeometric series}

Once we identify an appropriate $x_k$ for a physical system,
then we may use our general model to study the statistical
behavior of the system. 
The various models used in pion distribution can be related
to our general model with $\alpha_k = k$
by choosing $x_k$ as a term in a Gauss hypergeometric
series $F(a, b; c; z)$;
\begin{eqnarray}
 F(a, b; c; z) &=& \sum_{m=0}^\infty \frac{[a]_m [b]_m}{[c]_m} \frac{z^m}{m!}
                \label{hyperf}     \\
 {[a]_m} = \frac{\Gamma(a+m)}{\Gamma(a)}
   &=& \frac{\Gamma(a+n)}{\Gamma(a)} \frac{\Gamma(a+n + m-n)}{\Gamma(a+n)}
           \label{amfc}
   = [a]_n [a+n]_{m-n}       
\end{eqnarray}
Usefull values of $[a]_n$ are
 $[1/2]_n = (2n)!/(n! 2^{2n})$,
 $[1]_n = n!$,
 $[2]_n = (n+1)!$.
Considering only positive $k$, we choose
\begin{eqnarray}
 x_k &=& x \frac{[a]_{k-1} [b]_{k-1}}{[c]_{k-1}} \frac{z^k}{(k-1)!}
      = x \frac{\Gamma(a+k-1)}{\Gamma(a)} \frac{\Gamma(c)}{\Gamma(c+k-1)}
         \frac{\Gamma(b+k-1)}{\Gamma(b)} \frac{z^k}{(k-1)!}
\end{eqnarray}
For this case, the thermodynamic grand potential or the generating 
function is
\begin{eqnarray}
 f_0(\vec x) &=& f_0(x, z) = \log Z(x, z)
   = - \Omega(x, z) = \sum_{k=1}^\infty x_k  
    = x z F(a, b; c; z)        \label{f0hg}
\end{eqnarray}
If we allow jets without a pion, then we may allow $k = 0$ also.
For such a case,
\begin{eqnarray}
 x_k &=& x \frac{[a]_{k} [b]_{k}}{[c]_{k}} \frac{z^k}{k!}
      = x \frac{\Gamma(a+k)}{\Gamma(a)} \frac{\Gamma(c)}{\Gamma(c+k)}
         \frac{\Gamma(b+k)}{\Gamma(b)} \frac{z^k}{k!}   \\
 f_0(\vec x) &=& f_0(x, z) = \log Z(x, z)
   = - \Omega(x, z) = \sum_{k=0}^\infty x_k  
    = x F(a, b; c; z)
\end{eqnarray}
We can see the only difference of the generating functions
between the above two cases is the extra factor $z$ for
the grand potential.  
We will mostly concentrate on the first case, i.e., 
$k \ne 0$.

Using Eq. (\ref{factfm}) or the recurrence relation Eq.(\ref{recfm})
and Eqs.(\ref{hyperf}) -- (\ref{f0hg}),
\begin{eqnarray}
 f_m(x, z) &=& - z^m \left(\frac{d}{d z}\right)^m \Omega(x, z)
    = z^m \left(\frac{d}{d z}\right)^m \log Z(x, z)
    = z^m \left(\frac{d}{d z}\right)^m xz F(a,b;c;z)   \nonumber \\
   &=& x \frac{[a]_{m} [b]_{m}}{[c]_{m}} z^{m+1}
            F(a+m, b+m; c+m; z)
                \nonumber  \\   & &
     + x m \frac{[a]_{m-1} [b]_{m-1}}{[c]_{m-1}} z^{m}
            F(a+m-1, b+m-1; c+m-1; z)
\end{eqnarray}
The second order normalized factorial cumulant $\xi = \kappa_2$ and
the void variable $\chi$ are then
\begin{eqnarray}
 \xi(x, z) \bra A\ket &=& \frac{f_2(x, z)}{\bra A\ket}
    = \frac{f_2(x, z)}{f_1(x, z)}        \nonumber  \\
   &=& \frac{\frac{[a]_2 [b]_2}{[c]_2} z^2 F(a+2, b+2; c+2; z)
          + 2 \frac{a b}{c} z F(a+1, b+1; c+1; z)}
        {\frac{a b}{c} z F(a+1, b+1; c+1; z) + F(a, b; c; z)}     \\
 \chi(x, z) &=& \frac{f_0(x, z)}{\bra A\ket}
    = \frac{f_0(x, z)}{f_1(x, z)}     \nonumber  \\
   &=& \frac{F(a, b; c; z)}
         {\frac{a b}{c} z F(a+1, b+1; c+1; z) + F(a, b; c; z)}
\end{eqnarray}

For some values of $a$, $b$, and $c$, the hypergeometric
function become a simple function;
\begin{eqnarray}
 F(a, b; b; z) &=& (1-z)^{-a}   \nonumber  \\
 F(a, b; a; z) &=&(1-z)^{-b}            \nonumber   \\
 F(a, 1; 2; z) &=& \frac{1 - (1-z)^{1-a}}{z (1-a)}   \nonumber  \\
 F(1, 1; 2; z) &=& \lim_{a\to 1} F(a, 1; 2; z) = - \frac{\ln(1-z)}{z}
\end{eqnarray}
Various models of pion distributions can be related with
these functions as listed in Table \ref{tabl1}.

\begin{table}[htb]
\caption
{Various models with specific choice of $\alpha_k = k$ and 
$x_k$ in hypergeometric series $F(a, b; c; z)$ of Eq.(\protect\ref{hyperf}).
Here $k=0$ is not included, and
thus $f_0 = \ln Z = \sum_{k=1}^\infty x_k = xz F(a,b;c;z)$.
Here $1 \le k \le N$ 
with $N \to \infty$ except for Poisson which has a finite $Nx$.
Fisher exponent $\tau$ for each $x_k$ as discussed 
in Sect.\protect\ref{sectfrag} 
are given too.
 }
    \label{tabl1}
\begin{tabular}{c|c|c|ccc|c}
\hline
 Model                  &  $x_k$ & $f_0(\vec x) = \ln Z$ & $a$ & $b$ & $c$ & $\tau$ \\
\hline
 Poisson (P)            & $N x \delta_{k,1}$ or $x$ for $k = 1, 2, \cdots, N$
     &  $N x = \bar A$  &  &  &   &  \\
 Geometric (Geo)        &   $x z^k$   &  $\frac{xz}{1-z}$ & $a$  & 1 & $a$  & 0  \\
 Negative Binomial (NB) & $\frac{1}{k} x z^k$ & $-x \ln(1-z)$ & 1  & 1 & 2  & 1  \\
 Signal/Noise (SN)      & $(y + \frac{x}{k}) z^k$ 
       & $\frac{yz}{1-z} - x\ln(1-z)$  &  &  &   &  \\
 Lorentz/Catalan (LC) & $\frac{1}{k} 2^{-2(k-1)} \pmatrix{2(k-1) \cr k-1} xz^k$
       & $2x[1 - (1-z)^{1/2}]$  &  $\frac{1}{2}$   &  1  &  2   & 3/2  \\
 Hypergeometric (HGa)   & $\frac{[a]_{k-1}}{k!} x z^k$
      & $\frac{x}{1-a}[1 - (1-z)^{1-a}]$  &  $a$  &  1 &  2  & $2-a$ \\
 Random Walk--1d (RW1D) & ${2^{-2(k-1)}} \pmatrix{2(k-1) \cr k-1} x z^k$
      & $ x z (1-z)^{-1/2}$   &  $\frac{1}{2}$  & 1($b$)  & 1($b$)  &  1/2  \\
 Random Walk--2d (RW2D) 
   & $\left[{2^{-2(k-1)}} \pmatrix{2(k-1) \cr k-1}\right]^2 x z^k$
      &  $x z F(\frac{1}{2}, \frac{1}{2}; 1; z)$
        & $\frac{1}{2}$ &  $\frac{1}{2}$  & 1   &  1  \\
 Generalized RW1D (GRW1D)
   & $\frac{[a]_{k-1}}{(k-1)!} x z^k$ & $x z (1-z)^{-a}$ & $a$ & $b$ & $b$  &  $1-a$  \\
 Generalized RW2D (GRW2D)
   & $\left[\frac{[a]_{k-1}}{(k-1)!}\right]^2 x z^k$  & $x z F(a, a; 1; z)$
     & $a$ & $a$ & $1$   &  $2(1-a)$  \\
  \hline
\end{tabular} 
\end{table}

\begin{table}
\caption{ \protect
Factorial cumulants for various choices of $x_k$ of Table.\protect\ref{tabl1} 
  }  \label{tabl2}
\begin{tabular}{c|c|c|c|c}   
  \hline  
  Model  &  $f_0 = \log Z$  & $f_1 = \bra A\ket$  &  $z^{-2} f_2$  &
            $z^{-m} f_m$  \\
    \hline  
 P    & $N x = \bar A$  & $\bar A$  & 0 & 0 for $m \ge 2$ \\
 Geo  & $x \frac{z}{1-z}$  &  $x \frac{z}{(1-z)^2}$   &
      $x \frac{2}{(1-z)^3}$   &  $x \frac{m!}{(1-z)^{m+1}}$           \\
 NB & $- x \ln(1-z)$     &    $x \frac{z}{1-z}$    &
      $x \frac{1}{(1-z)^2}$   &  $x \frac{(m-1)!}{(1-z)^m}$           \\
 SN   & $\frac{yz}{1-z} -x \ln(1-z)$ & $y\frac{z}{(1-z)^2} + x\frac{z}{1-z}$ &
      $y \frac{2}{(1-z)^3} + x \frac{1}{(1-z)^2}$   &
       $\frac{(m-1)!}{(1-z)^m} \left(y \frac{m}{(1-z)} + x\right)$  \\
 LC   & $2 x [1 - (1-z)^{1/2}]$  & $x \frac{z}{(1-z)^{1/2}}$  &
   $x \frac{1/2}{(1-z)^{3/2}}$  &  $x \frac{[1/2]_m}{(1-z)^{m-1/2}}$   \\
 HGa  & $\frac{x}{1-a} [1 - (1-z)^{1-a}]$ &
    $x \frac{z}{(1-z)^a}$  &  $x \frac{a}{(1-z)^{a+1}}$   &
            $x \frac{[a]_{m-1}}{(1-z)^{a+m-1}}$           \\
 RW1D  & $x \frac{z}{(1-z)^{1/2}}$   &
    $\frac{x}{2} \frac{z}{(1-z)^{1/2}} + \frac{x}{2} \frac{z}{(1-z)^{3/2}}$
  & $\frac{x}{4} \frac{1}{(1-z)^{3/2}} + \frac{3}{4} x \frac{1}{(1-z)^{5/2}}$
  & $- x \frac{[-1/2]_m}{(1-z)^{a+1}} + x \frac{[1/2]_m}{(1-z)^{a+2}}$
         \\
 GRW1D   & $x \frac{z}{(1-z)^{a}}$
  & $- x \frac{(a-1) z}{(1-z)^{a}} + {x} \frac{a z}{(1-z)^{a+1}}$
  & $- x \frac{(a-1) a}{(1-z)^{a+1}} + x \frac{a (a+1)}{(1-z)^{a+2}}$
  & $- x \frac{[a-1]_m}{(1-z)^{a+m-1}} + x \frac{[a]_m}{(1-z)^{a+m}}$
         \\
    \hline  
\end{tabular}
\end{table}

More detail discussions and related physical systems  
of these distributions will be given and discussed in the next section.
Brief discussion about the weight $x_k$ of each model follows.
All the distributions, except the Poisson distribution (P), 
considered in Table \ref{tabl1} 
have several factors in the weight $x_k$.
One factor in $x_k$ is $z^k$ which is a $k$ dependent geometric term
and comes from assigning the same weight $z$ to each constituents
independent of the cluster or cycle classes it belongs to.
Another factor of weight is independent of $k$ such as the $x$ 
in Table \ref{tabl1} which comes from assigning the same weight $x$
to each cluster or the cycle class as a whole 
independent of its internal structure.
These two factors, $x z^k$, are multiplied by a $k$ dependent or 
independent prefactor.
A geometric (Geo) distribution follows when there is no other weight 
factor beside $x$ and $z$, i.e., no $k$ dependent prefactor 
so that $x_k = x z^k$.
The Geo with $z = 1$ for a finite $N$ and with $z = 0$ for $k > N$ is 
the same as the Poisson distribution; both have $f_0 = N x$.
The negative binomial (NB) which appears frequently in various
studies has a weight factor assigned to a cluster or cycle
class given by $x z^k/k$.
This has an extra size dependent factor of $1/k$
compared to the geometric distribution.
The signal/noise model (SN) has a two part structure and
interpolates between a Poisson and NB distribution.
The geometric distribution is the signal component of SN
while the NB distribution is the noise component of SN.
The Lorentz/Catalan model (LC) has in its weight a shifted Catalan number
divided by $2^{2(k-1)}$, that is  
$\frac{[1/2]_{k-1}}{k!} = \frac{2^{-2(k-1)}}{k} \pmatrix{2(k-1) \cr k-1}$,
beside the $x z^k$ factor which is the weight for Geo model.
The Catalan numbers given by
 $\pmatrix{2k \cr k}/(k+1)$ are 1, 2, 5, 14, $\cdots$ 
for $k = 1$, 2, 3, 4, $\cdots$ and 
the shifted Catalan numbers given by 
 $\pmatrix{2(k-1) \cr k-1}/k$ are 1, 1, 2, 5, 14, $\cdots$.
The importance of this factor is shown in Fig. \ref{LCfig} 
of section \ref{ancest}.  

 Of all these distributions, the NB has been the most
frequently studied. Ref.\cite{poistrn}
gives several sources for its origin. These sources include
sequential processes, self-similar cascade models and connections with
Cantor sets and fractal stucture, generalizations of the Planck
distribution, solutions to stochastic differential equations. 
Becattini \cite{becat}
have shown that the NB distribution arises from decaying resonances. 
The $\alpha$ model of Ref.\cite{ref9}, which
is a self similar random cascade process, leads to a NB like behavior. 
The stochastic aspects of the NB distribution have been discussed 
by R. Hwa \cite{rhwa}. Hegyi \cite{hegyi} has discussed
the NB distribution in terms of combinants. As already mentioned, the LC
model can be connected to a Ginzburg-Landau approach and also has an
underlying splitting or branching dynamics and cascade like features.

As can be seen from the arguments of the hypergeometric function
$F(a, b; c; z)$ in Table \ref{tabl1}, the hypergeometric model 
with $b = 1$ and $c = 2$ (HGa) 
include Geo, NB, SN, LC as a special case of HGa 
depending on the value of $a$.
Other models listed in Table \ref{tabl1} are based on random walks.
The use of random walk results was originally due to Feynman \cite{feyn}
in his description of the phase transition in liquid helium.
The random walk aspects arise when considering the closing of cycle
of length $k$. We include them for completeness.
Since the random walk in 1-dimension (RW1D) is the same as LC
except the missing $1/k$ dependence compared to LC,
RW1D can be extended to a generalized RW1D (GRW1D) similar to
the generalization of LC to HGa.
A random walk model in 2-dimension has an extra factor of a shifted
Catalan number and $k 2^{-2(k-1)}$ factor
compared to RW1D and can also be generalized to GRW2D.

Since $k = 0$ is excluded here, 
the partition function for these models are
given simply by a hypergeometric function 
as $Z = \exp[\sum_{k=1}^\infty x_k] = \exp[xz F(a,b;c;z)]$
with various choices of $a$, $b$, $c$.
For example the LC model has $f_0 = \sum_k x_k = x z F(1/2,1;2;z)$
and the NB model has $f_0 = xz F(1,1;2;z)$.
The geometric model $x_k = y z^k$ has
 $f_0 = yz F(a,1;a;z) = yz F(2,1;2;z)$
while the SN model is a combination of the geometric plus NB cases.
These functions are special cases of $f_0 = x z F(a, 1; 2; z)$ of 
the HGa model. 
The generalized random work in 1-dimension (GRW1D) has
$f_0 = xz F(a, b; b; z)$ and
the generaized RW in 2-dimension (GRW2D) has
$f_0 = xz F(a,a; 1; z)$.
The factorial cumulants $f_m$ for these models are summarized
in Table \ref{tabl2}.
Several cases with $c = 2$ have a canonical partition function $Z_n$ 
which can be writen in terms of
confluent hypergeometric functions $U(u,v;w)$
and standard factor $z^n/n!$ \cite{prc65}.

\subsection{Generalized model of Hypergeometric (HGa)}  \label{glcbas}

We consider in more detail the hypergeometric model with $b=1$ and $c=2$ (HGa)
here since it includes the NB, Geo, and LC models as special cases.
This generalized model is related with
the hypergeometric function with $b=1$ and $c=2$ 
with an arbitrary value of $a$, 
i.e., $F(a,1;2;z)$, and has the weight of
\begin{eqnarray}
 x_k &=& x z^k \frac{[a]_{k-1}}{k!} = x \frac{z^k}{k!} \frac{\Gamma(a+k-1)}{\Gamma(a)}
\end{eqnarray}
The asymptotic behavior of $x_k$ is
\begin{eqnarray}
   x_k &=& x z^k k^{a-2}/\Gamma(a) = x z^k k^{-\tau}/\Gamma(a)
\end{eqnarray}
for large $k$ using Stirling approximation. 
Thus
\begin{eqnarray}
   \tau &=& 2-a 
\end{eqnarray}
which connects the parameter $a$ to the physical Fisher critical exponent
$\tau$ for the HGa class of distributions.
Its associated grand canonical partition function  
\begin{eqnarray}
 Z(x, z) &=& e^{f_0} = e^{xz F(a,1;2;z)}
    = \exp\left[\frac{x}{(a-1)} \left(\frac{1}{(1-z)^{(a-1)}} - 1\right)\right]    
\end{eqnarray}
is shown in Table \ref{tabl1}.
From Table \ref{tabl2}, we have
\begin{eqnarray}
 f_m(x, z) &=& [a]_{m-1} \frac{x z^m}{(1-z)^{a+m-1}} 
    = x \frac{\Gamma(a+m-1)}{\Gamma(a)} \frac{z^m}{(1-z)^{a+m-1}}  \\
 \kappa_m(x, z) &=& \frac{[a]_{m-1}}{x^{m-1}} (1-z)^{(a-1)(m-1)}
    = \frac{\Gamma(a+m-1)}{\Gamma(a)}
            \left(\frac{(1-z)^{(a-1)}}{x}\right)^{m-1}        \nonumber  \\
  &=& \frac{\Gamma(a + m - 1)}{\Gamma(a)} \frac{\kappa_2^{m-1}(x,z)}{a^{m-1}}
     = A_m \kappa_2^{m-1}(x,z)                \label{kappan}
\end{eqnarray}
The normalized factorial cumulant, i.e., 
the reduced cumulant $\kappa_m$ shows the hierarchical structure
of HGa at the reduced cumulant level 
with $A_m = a^{-(m-1)} \Gamma(a+m-1)/\Gamma(a)$
where $\kappa_m$ is related to $\kappa_2$.
This property was realized for the NB distribution in Ref.\cite{hiera}
which is obtained for Eq.(\ref{kappan}) with $a=1$ giving $A_m = (m-1)!$.
The result of Eq.(\ref{kappan}) is a generalization of the NB result.

Some moments for HGa are
\begin{eqnarray}
 \bra A\ket(x, z) &=& f_1(z, \vec x) = \frac{x z}{(1-z)^a}      \\
 \chi(x, z) &=&\frac{f_0(x,z)}{f_1(x,z)}
     = \frac{1}{(1-a)} \frac{(1-z)}{z} \left[(1-z)^{a-1} - 1\right]       \\
 \xi(x, z) &=& \kappa_2(x, z) = \frac{f_2(x,z)}{f_1^2(x,z)} = \frac{a}{x} (1-z)^{(a-1)}  
\end{eqnarray}
Since these relations give
\begin{eqnarray}
 \xi(x, z) \bra A\ket(x, z) &=& \kappa_2(x, z) \bra A\ket(x, z) 
    =  \frac{f_2(x, z)}{f_1(x, z)}  
    = \frac{a z}{(1-z)} 
\end{eqnarray}
the void parameters can be obtained in terms of the normalized 
fluctuation $\xi$ and the mean number $\bra A\ket = \bar A$ by
\begin{eqnarray}
 z(\bar A, \xi) &=& \frac{\xi\bar A}{a + \xi\bar A} = \frac{f_2/\bar A}{a + f_2/\bar A}
    = \frac{f_2}{f_2 + a \bar A}          \\
 x(\bar A, \xi) &=& \frac{\bar A}{z} (1-z)^a
    = \frac{a}{\xi} \left(\frac{a}{a + \xi\bar A}\right)^{(a-1)}
    = \frac{a \bar A^2}{f_2} \left(\frac{a \bar A}{a \bar A + f_2}\right)^{a-1}
                 \label{xmnxiglc}  \\
 f_0(\bar A, \xi) &=& \log Z(\bar A, \xi) 
  = \frac{x}{a-1} \left[\left(1 + \frac{\xi\bar A}{a}\right)^{a-1} - 1\right]
 = \frac{x}{a-1} \left[\left(1 + \frac{f_2}{a\bar A}\right)^{a-1} - 1\right] \\
 \chi(\bar A, \xi) &=& \frac{f_0}{\bar A}
  =  \frac{1}{(1-a)} \frac{a}{\xi \bar A}
       \left[\left(1 + \frac{\xi \bar A}{a}\right)^{1-a} - 1\right]
   =  \frac{1}{(1-a)} \frac{a \bar A}{f_2}
        \left[\left(1 + \frac{f_2}{a \bar A}\right)^{1-a} - 1\right]
               \label{chiglc}   \\
 \kappa_m(\bar A, \xi) &=& \frac{f_m(\bar A, \xi)}{\bar A ^m}
  = \frac{\Gamma(a+m-1)}{\Gamma(a)} \left(\frac{\xi}{a}\right)^{m-1}
              \label{kapanglc}
\end{eqnarray}
for a given mean value of $\bra A\ket = \bar A$ and 
the fluctuation $\xi$ or $\xi\bar A$ or $f_2$.
Using these $x(\bar A, \xi)$ and $z(\bar A, \xi)$ 
we can easily find the multiparticle distribution $P_A(\bar A, \xi)$ for
a given values of mean $\bar A$ and fluctuation $\xi$ 
using the recurence relation of Eq.(\ref{recur}) for $Z_A(\bar A, \xi)$.
We explicitly compare the distribution $P_A$ for different values of $a$ 
with given values of $\bar A$ and $\xi$ in Sect.\ref{sectcomp}.

Table \ref{tabl1} shows that the generalized HGa model becomes 
the Lorentz/Catalan (LC) model with $a = 1/2$,
the negative binomial (NB) model with $a = 1$, 
and geometric (Geo) distribution with $a = 2$.
However the NB should be considered as a $a \to 1$ limit of HGa;  
\begin{eqnarray}
 \lim_{a \to 1} f_0 &=& -x \log (1-z)        \\
 \lim_{a \to 1} Z &=& (1-z)^{-x}             \\
 \lim_{a\to 1} f_0(\bar A, \xi) &=& x \ln \left(1 + \xi\bar A\right)
    = x \ln \left(1 + f_2/\bar A\right)          \\
 \lim_{a \to 1} \chi(\bar A, \xi)
   &=& \frac{1}{\xi \bar A} \log\left(1 + \xi \bar A\right) 
\end{eqnarray}
Thus a NB distribution has
 $\chi = \ln(1 + \xi \bar A)/(\xi \bar A)$ with $x = 1/\xi$
while the LC distribution has $\chi = (\sqrt{1+2\xi \bar A} - 1)/(\xi \bar A)$
with $x = \sqrt{1 + 2 \xi \bar A}/(2\xi)$
and a Geo distribution has $\chi = 2 [1 - (1+\xi \bar A/2)^{-1}]/(\xi \bar A)$
with $1/x = \xi(1 + \xi \bar A/2)/2$
as limiting expressions of a more general $\chi$ given by Eq.(\ref{chiglc}) 
for HGa.
Eq.(\ref{xmnxiglc}) shows that the fluctuation $\xi$ depends on 
the mean $\bar A$ for given $x$ except for NB. 
Since $\xi \ge 0$ for $\alpha_k = k$ with non-negative $x_k$,
$\xi(\bar A, x) = 1/x$ for $a=1$ (NB),
$\xi(\bar A, x) = [\bar A + \sqrt{\bar A^2 + 4x^2}]/(4x^2)$ for $a=1/2$ (LC),
$\xi(\bar A, x) = [-x + \sqrt{x^2 + 4x\bar A}]/(x\bar A)$ for $a=2$ (Geo).
However for small $\bar A$ and large $x$ the fluctuation $\xi$ is
inependent of the mean $\bar A$ and $\xi(\bar A, x) \approx a/x$ for HGa.

The void parameters are $\xi = 0$ and $\chi = 1$ for Poisson distribution.
The reduced cumulants are $\kappa_m = 0$ for $m > 2$ for Poisson distribution.
The void parameters are
 $\xi = \frac{1}{x}$ and $\chi = \frac{\ln(1 + \xi \bar A)}{\xi \bar A}$
for NB distribution.
The $m$-th order reduced cumulant is $\kappa_m = (m-1)! \xi^{m-1}$ for NB.
For SN distribution, the void parameters are
\begin{eqnarray}
 \xi &=& \frac{f_2}{f_1^2} = \frac{N}{x} \frac{(2S+N)}{(S+N)^2}     \\
 \chi &=& \frac{f_0}{f_1} = \frac{ x \ln(1+N/x) + S/(1+N/x)}{S+N}
\end{eqnarray}
with signal level $S = \frac{yz}{(1-z)^2}$ and noise level $N = \frac{xz}{1-z}$.
The SN model has important application to quantum optics
and, in particluar, to photon counts from lasers \cite{qoptic}.
Biyajima \cite{biyaj} has suggested using it for particle
multiplicity distribution as does Ref. \cite{poistrn}.
When the noise level $N \to 0$, $\xi \to 0$ and $\chi \to 1$.
When the signal level $S \to 0$, $\xi \to 1/x$ 
and $\chi \to \frac{x}{N} \ln(1 + N/x)$.
The third order reduced cumulant is
 $\kappa_3 = \frac{2!}{(1-z)^3} \left(\frac{3y}{(1-z)} + x\right)
    = 2 \frac{x}{N} (1 + N/x)^3 (3S + N)$ for SN.
For LC distribution,
the void parameters are  $\xi = \frac{1}{2x \sqrt{1-z}}$
and $\chi = \frac{1}{\xi \bar A}\left[\sqrt{1 + 2 \xi \bar A} - 1\right]$
with $z = 2 \xi \bar A /(1 + 2 \xi \bar A)$.
The $m$-th order reduced cumulant is
 $\kappa_m = [\frac{1}{2}]_{m-1} (2 \xi)^{m-1}$ for LC.
The results of quantum optics, in the notation of Ref. \cite{qoptic},
can be obtained \cite{prl86} when $x = T \Omega/2$,
 $z = 2 W \gamma/\Omega^2$, $\Omega^2 = \gamma^2 + 2 W \gamma$.
The $W$ is an integral of the Lorentzian line shape
 $\Gamma(\omega) = b/[(\omega - \omega_0)^2 + \gamma^2]$,
 $T$ is the time, and $2 x \sqrt{1-z} = \gamma T$.

\subsection{Generalized random walk in one dimension }

The hypergeometric model with $c=b$ has $x_k = x \frac{[a]_{k-1}}{(k-1)!} z^k$
which is a generalized random walk process in one dimension (GRW1D).
For $a=1/2$ the GRW1D becomes the random walk in one dimension.
Since $F(a, b; c; z) = F(b, a; c; z)$, it is easy to see that
both HGa with $a=2$ and GRW1D with $a=1$ are geometric models.
For GRW1D, we have $f_0 = x \frac{z}{(1-z)^a}$ which is $f_1$ of HGa
as shown in Table II.
The asymptotic behavior of $x_k = x \frac{[a]_{k-1}}{(k-1)!} z^k$ is
 $x_k = x z^k k^{a-1}/\Gamma(a)$
for large $k$ and thus 
the Fisher critical exponent is $\tau = 1-a$ for the GRW1D class of distributions.
Its associated grand canonical partition function is
\begin{eqnarray}
 Z(x, z) &=& e^{f_0} = e^{xz F(a,b;b;z)} = \exp\left[\frac{x z}{(1-z)^a}\right]
\end{eqnarray}
The factorial cumurants for GRW1D are, from Table II, 
\begin{eqnarray}
 f_m(x, z) &=& - x \frac{[a-1]_m}{(1-z)^{a+m-1}} z^m
              + x \frac{[a]_m}{(1-z)^{a+m}} z^m
            \nonumber  \\
  &=& x [a]_{m-1} \frac{\left[(a-1)z + m\right]}{(1-z)^{a+m}} z^m   \\
 \kappa_m(x, z) 
  &=& x \frac{[a]_{m-1} z^m}{(1-z)^{a+m}} \left[(a-1)z + m\right]  
     \left[x \frac{z}{(1-z)^{a+1}}\right]^{-m} \left[(a-1)z + 1\right]^{-m}
            \nonumber  \\
  &=& \left[\frac{(1-z)^a}{x \left[(a-1)z + 1\right]}\right]^{m-1} 
        \frac{[a]_{m-1} \left[(a-1)z + m\right]}{\left[(a-1)z + 1\right]}
            \nonumber  \\
  &=& [a+1]_{m-2} \frac{\left[(a-1)z + m\right]}{\left[(a-1)z + 2\right]}
       \left[\frac{[(a-1)z + 1]}{a [(a-1)z + 2]}\right]^{m-2} \kappa_2^{m-1}
   = A_m \kappa_2^{m-1}    \label{kapmgrw}
\end{eqnarray}
Here the coefficient $A_m$ that appears in the hierarchical structure relation 
of Eq.(\ref{kapmgrw}) depends also on $z$ 
which means that $A_m$ depends on the mean $\bar A$ and the normalized
fluctuation $\xi$. For HGa $A_m$ is independent of $\bar A$ and $\xi$.

Some moments for GRW1D are
\begin{eqnarray}
 \bra A\ket(x,z) &=& \bar A = f_1(x,z)
      =  \frac{x z}{(1-z)^{a+1}} [(a-1)z + 1]    \\
 \chi(x,z) &=& \frac{f_0(x,z)}{f_1(x,z)}
      = \frac{(1-z)}{[(a-1)z + 1]}    \\
 \xi(x,z) &=& \kappa_2(x,z) = \frac{f_2(x,z)}{f_1^2(x,z)}
     =  \frac{a}{x} (1-z)^a \frac{[(a-1)z + 2]}{[(a-1)z + 1]^2}
\end{eqnarray}
Since these gives
\begin{eqnarray}
 \xi(x,z) \bra A\ket(x,z) &=& \kappa_2(x,z) \bra A\ket(x,z) = \frac{f_2(x,z)}{f_1(x,z)}
      = \frac{a z}{(1-z)} \frac{[(a-1)z + 2]}{[(a-1)z + 1]}
\end{eqnarray}
we have,
for a given mean value $\bra A\ket = \bar A$ and the fluctuation $\xi$ or $f_2$,
\begin{eqnarray}
 z(\bar A, \xi) &=& \frac{(a-2) \xi \bar A - 2 a 
                       \pm a \sqrt{(\xi \bar A)^2 + 4 \xi \bar A/a + 4}}
                      {2 (a-1) (\xi \bar A + a)}
     ~~\buildrel\Longrightarrow\over{_{a\to 1}}~~
            \frac{\xi\bar A}{2 + \xi\bar A}  \\
 x(\bar A, \xi)
  &=& \frac{\bar A}{z} \frac{(1-z)^{a+1}}{[(a-1)z + 1]} 
               \nonumber  \\
  &=& \frac{\bar A}{z}
       \left(\frac{2(\xi \bar A + a)}{a \xi \bar A  
                   \pm a \sqrt{(\xi \bar A)^2 + 4 \xi \bar A/a + 4}}\right)
       \left(\frac{a \xi \bar A + 2 a^2 
                       \mp a \sqrt{(\xi \bar A)^2 + 4 \xi \bar A/a + 4}}
                      {2 (a-1) (\xi \bar A + a)}\right)^{a+1}
                   \label{xmnxigrw}    \\
 f_0(\bar A, \xi) &=& x \frac{z}{(1-z)^a}
     = x z \left(\frac{2 (a-1) (\xi \bar A + a)}{a \xi \bar A + 2 a^2 
             \mp a \sqrt{(\xi \bar A)^2 + 4 \xi \bar A/a + 4}} \right)^{a}  \\
 \chi(\bar A, \xi) &=& \frac{f_0}{\bar A}
   = \left(\frac{\xi \bar A + 2 a
                       \mp \sqrt{(\xi \bar A)^2 + 4 \xi \bar A/a + 4}}
            {\xi \bar A  
                   \pm \sqrt{(\xi \bar A)^2 + 4 \xi \bar A/a + 4}}\right)
        \frac{1}{(a-1)}
    ~~\buildrel\Longrightarrow\over{_{a\to 1}}~~ \frac{2}{2 + \xi\bar A}  \\
 \kappa_m(\bar A, \xi) &=& \frac{f_m(\bar A, \xi)}{\bar A^m}
   = [a]_{m-1} \left[\frac{(a-1)z + m}{(a-1)z + 2}\right]
       \left[\frac{(a-1)z + 1}{(a-1)z + 2}\right]^{m-2}
       \left(\frac{\kappa_2}{a}\right)^{m-1}
               \nonumber  \\
   &=& \left(\frac{(a + 2(m - 1)) \xi \bar A + 2(m - 1) a 
                \pm a \sqrt{(\xi \bar A)^2 + 4 \xi \bar A/a + 4}}
                    {(a + 2) \xi \bar A + 2 a 
                \pm a \sqrt{(\xi \bar A)^2 + 4 \xi \bar A/a + 4}}\right)
               \nonumber  \\
     & & \ \ \times 
         \left(\frac{a \xi \bar A 
                \pm a \sqrt{(\xi \bar A)^2 + 4 \xi \bar A/a + 4}}
                    {(a + 2) \xi \bar A + 2 a 
                \pm a \sqrt{(\xi \bar A)^2 + 4 \xi \bar A/a + 4}}\right)^{m-2}
           [a]_{m-1} \left(\frac{\xi}{a}\right)^{m-1}
\end{eqnarray}
Here the condition $0 < z < 1$ determines the right one of $\pm$ sign.
Note here that
 $\kappa_m(\bar A, \xi) = [a]_{m-1} \left(\frac{\xi}{a}\right)^{m-1}$
for HGa.
For the GRW1D, the normalized factorial cumulant $\kappa_m(\bar A, \xi)$ 
has an extra dependence of the order $m$ which also depends 
on the mean $\bar A$ and the fluctuation $\xi$.

\section{Comparison of Probability distributions within HGa}
      \label{sectcomp}

The hypergeometric  
case (HGa) which we studied in more detail in Sect. \ref{glcbas}
included various cases of Table \ref{tabl1}; Geo, NB, LC
which are distinguished by one parameter $a$ of the hypergeometric model.
Thus we use HGa to compare various models for pion distribution
and other particle distributions, with particle number $n = \sum_k k n_k$.

\subsection{Voids and void scaling relations}  \label{voidscal}

\begin{figure}[htb]
\centerline{
   \epsfxsize=4.0in   \epsfbox{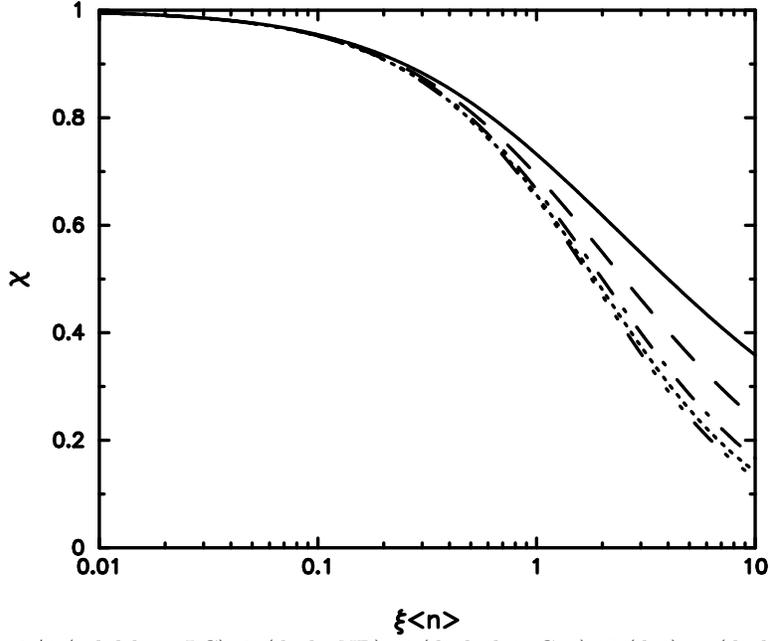}
}
\caption{$\chi$ vs $\xi\bra n\ket$ for $a = 1/2$ (solid line; LC), 
1 (dash; NB), 2 (dash-dot; Geo), 3 (dot), 4 (dash-dot-dot-dot).
For Poisson distribution $\xi = 0$ and $\chi=1$.}
 \label{hierafig}
\end{figure}

Void analysis looks for scaling propoerties associated with $\chi = f_0/\bra n\ket$;
specifically, $\chi$ is a function of the combination $\xi\bra n\ket = f_2/\bra n\ket$
where $\xi = \kappa_2$ is the coefficient of $\bra n\ket^2$ in the fluctuation
 $\sigma^2 = \bra n^2\ket - \bra n\ket^2 = \bra n\ket + \xi \bra n\ket^2$.
For generalized HGa model $\chi$ is given by
\begin{eqnarray}
 \chi = \frac{\left(1 + \xi\bra n\ket/a\right)^{1-a} - 1}{(1-a)\xi\bra n\ket/a}
\end{eqnarray}
We show the void variable $\chi$ as a function of $\xi \bra n\ket$ in Fig.\ref{hierafig}.
This shows that we can vary $a$ to fit data.
Ref.\cite{hiera} claims the NB distribution (HGa with $a=1$) fits reasonably
well the void distribution for single jet events in $e^+ e^-$ annihilation 
but Fig.\ref{hierafig} shows 
all of the curves might fit such data up to $\xi\bra n\ket \sim 1$
since the various curves are reasonably close up to this $\xi\bra n\ket$.
Thus further investigations of this data is required to distinguish
various models. 
Higher moments or cumulants might need to be compared for this purpose.
Within HGa, $\kappa_3(\bar n, \xi, a) = \left(\frac{a+1}{a}\right) \xi^2$ 
for a given value of $\bra n\ket = \bar n$ and $\xi$ 
according to Eq.(\ref{kapanglc}).
Thus $A_3 = \kappa_3/\xi^2 = (a+1)/a = 3$, 2, 3/2, 4/3, 5/4 
for $a = 1/2$, 1, 2, 3, 4 independent of $\xi$.
The $\kappa_3$ may help in distinguishing various models
for the data with a small value of $\xi$, i.e., the region of $\xi \bar n < 1$
in Fig.\ref{hierafig}.
The differences of $A_m$ between models with different value of $a$ 
becomes larger as the order $m$ of the reduced factorial cummulant
becomes higher.

Table \ref{tabldat} shows the values of $\xi$ and $\kappa_3$ 
of charged particle multiplicity distribution of jets
for L3 data \cite{jeteel3} and H1 data \cite{jeteph1}.
From these values $A_3 = 4.1686$ and $-6.2653$ for L3 and H1 respectively.
The corresponding values of HGa model parameter $a$ are 
0.31559 and $-0.13764$ respectively
and the corresponding values of GRW1D model parameter $a$ are 
0.14006 and $-0.8516$. 
The negative value of $a$ makes the corresponding $z$ value in HGa larger 
than 1 which is not allowed because of Eq.(\ref{xmnxiglc}) for real $x$.
Thus the HGa model cannot fit H1 data 
with the same values of mean, fluctuation and $\kappa_3$ at the same time.
The negative value of $\kappa_3$ or $A_3$ causes the negative values of
$a$, $x$ and $z$ in GRW1D which are acceptable in Eq.(\ref{xmnxigrw}) for real $x$. 
This model with $a = - 0.8516$ fits H1 data best for all the values
of mean, fluctuation, $\kappa_3$, and $\chi = f_0/f_1$.
The Fisher exponent for this parameter is $\tau = 1-a = 1.85$.
The physical meaning of the negative $x$ and $z$ should be studied further 
similar as studied in Ref.\cite{frag}.
Both HGa and GRW1D models with small values of $a$ (0.31559 and 0.14006 
respectively) fit the L3 data well with the same mean, fluctuation, 
and the third cumulant at the same time.
The corresponding values of Fisher exponents are $\tau = 2-a = 1.68$
and $\tau = 1-a = 0.86$ respectively.
If we can extract the value of $\tau$ from data then we may distinguish which model
fits the data better.

\begin{table}
\caption{Various cumulants of charged particle multiplicity distribution
of jets for L3 data (all events of $e^+e^-$) \protect\cite{jeteel3} 
and H1 data (pseudorapidity range $1 < \eta^* < 5$ with 80 GeV $< W <$ 115 GeV
of $e^+p$) \protect\cite{jeteph1} compared with
models having the same mean and fluctuation.
}  \label{tabldat}
\begin{tabular}{lccccccc}
 \ model &   $a$   &  $x$   &  $z$  & $\bra n\ket$ & $\xi$  &$\kappa_3$&$\chi$  \\
\hline
 L3 data \ &       &        &         & 20.463 & 0.044238 & 0.008158 &       \\
 HGa     & \ 0.5 (LC) \ & \ 18.948 \ & \ 0.64419 \ & \ 20.463 \ 
             & \ 0.044238 \ & \ 0.005871 \ & \ 0.74726 \ \\
         & 1.0 (NB)& 22.605 & 0.47514 & 20.464 & 0.044238 & 0.003914 & 0.71208 \\
         &2.0 (Geo)& 31.123 & 0.31159 & 20.463 & 0.044238 & 0.002936 & 0.68841 \\
         & 0.31559 & 18.007 & 0.74150 & 20.463 & 0.044238 & 0.008158 & 0.77636 \\
 GRW1D   & 0.14006 & 20.905 & 0.66088 & 20.463 & 0.044239 & 0.008158 & 0.78558 \\
\hline
 H1 data &         &        &         & 7.7210 & 0.069186 &--0.000764& 0.73242 \\
 HGa     & 0.5 (LC)& 10.394 & 0.51653 & 7.7210 & 0.069186 & 0.014360 & 0.82028 \\
         & 1.0 (NB)& 14.454 & 0.34819 & 7.7210 & 0.069186 & 0.009573 & 0.80122 \\
         &2.0 (Geo)& 22.814 & 0.21079 & 7.7210 & 0.069186 & 0.007180 & 0.78921 \\
         &    3.0  & 31.244 & 0.15115 & 7.7210 & 0.069186 & 0.006382 & 0.78470 \\
 GRW1D   &--0.85160&--4.4150&--0.81168& 7.7211 & 0.069185 &--0.000764& 0.72375 \\
\end{tabular}
\end{table}

\subsection{Probability distribution}

\begin{figure}[htb]
\centerline{
   \epsfxsize=5.5in   \epsfbox{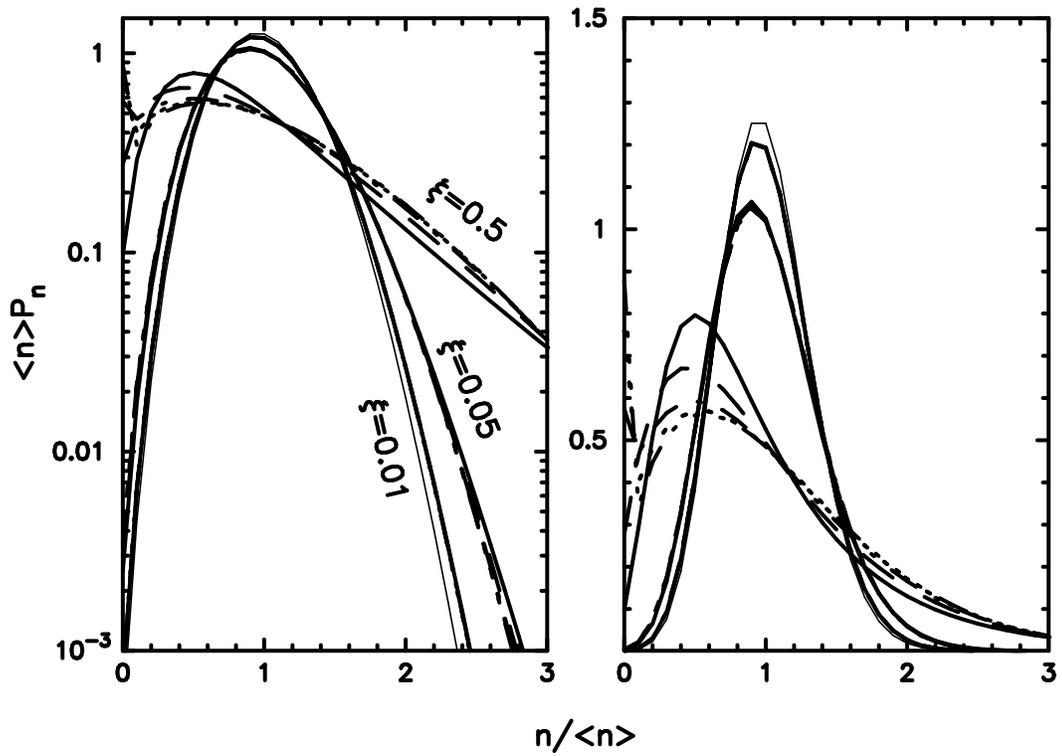}
}
\caption{$P_n$ for fixed $\bra n\ket = 10$ for $a = 1/2$ (solid line), 1 (dash), 
2 (dash-dot), 3 (dot), 4 (dash-dot-dot-dot)
and for $\xi = 0.01$, 0.05, and 0.5
in log scale on the left and linear scale on the right.
For $\xi = 0.01$ all distributions become
very close to Poisson (thin solid curve) already.
($P_0$ becomes large for large $\xi$.)}
 \label{cankno}
\end{figure}

\begin{figure}[htb]
\centerline{
   \epsfxsize=5.5in   \epsfbox{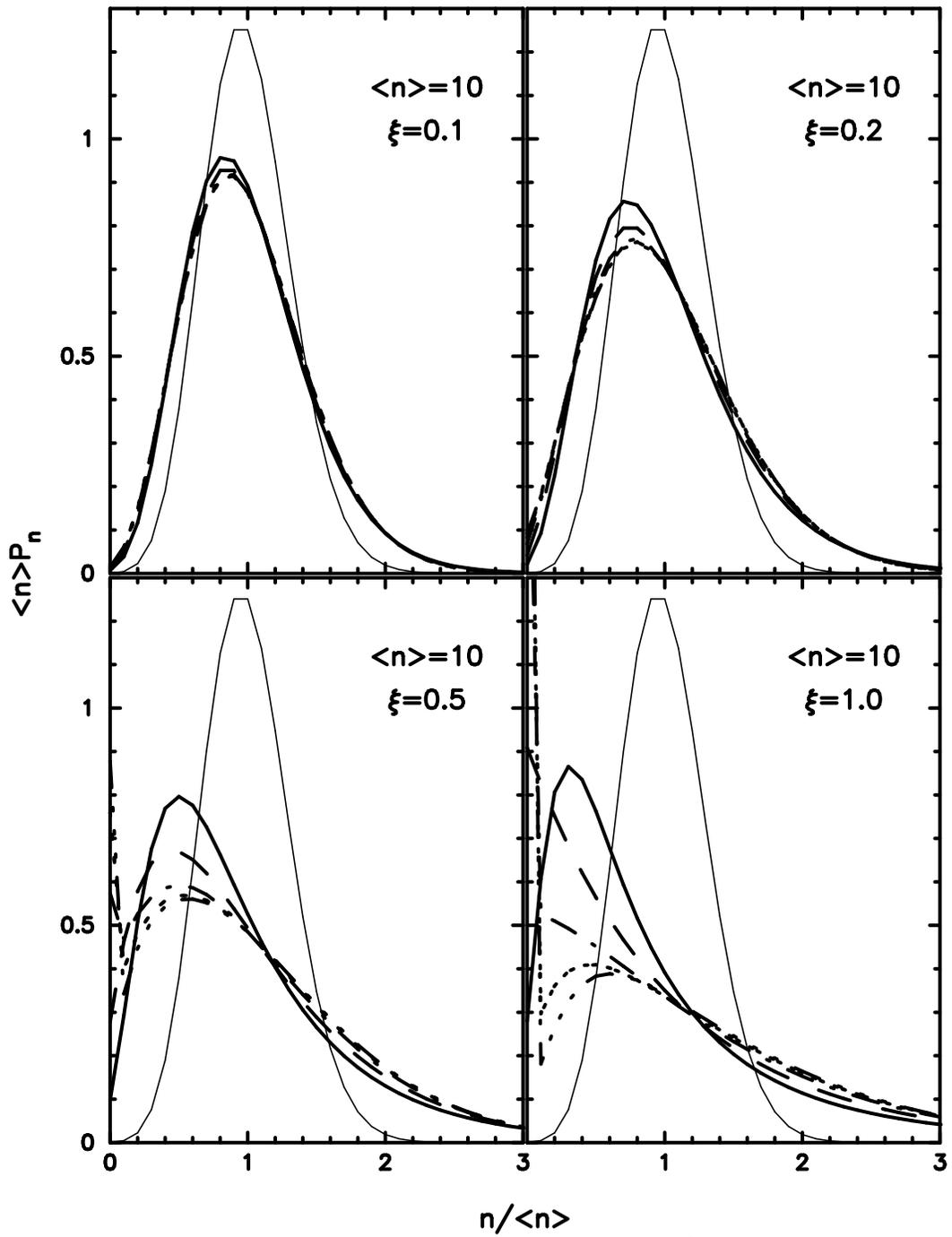}
}
\caption{ 
$P_n$ for fixed $\bra n\ket = 10$ and for $a = 1/2$, 1, 2, 3, 4. 
The value $\xi = f_2/\bra n\ket^2$ are shown in each figure.
The various choices of $a$ for each curve are given in the figure
caption of Figs. \protect \ref{hierafig} and \protect \ref{cankno}.   }
 \label{canpna}
\end{figure}

\begin{figure}[htb]
\centerline{
   \epsfxsize=5.5in   \epsfbox{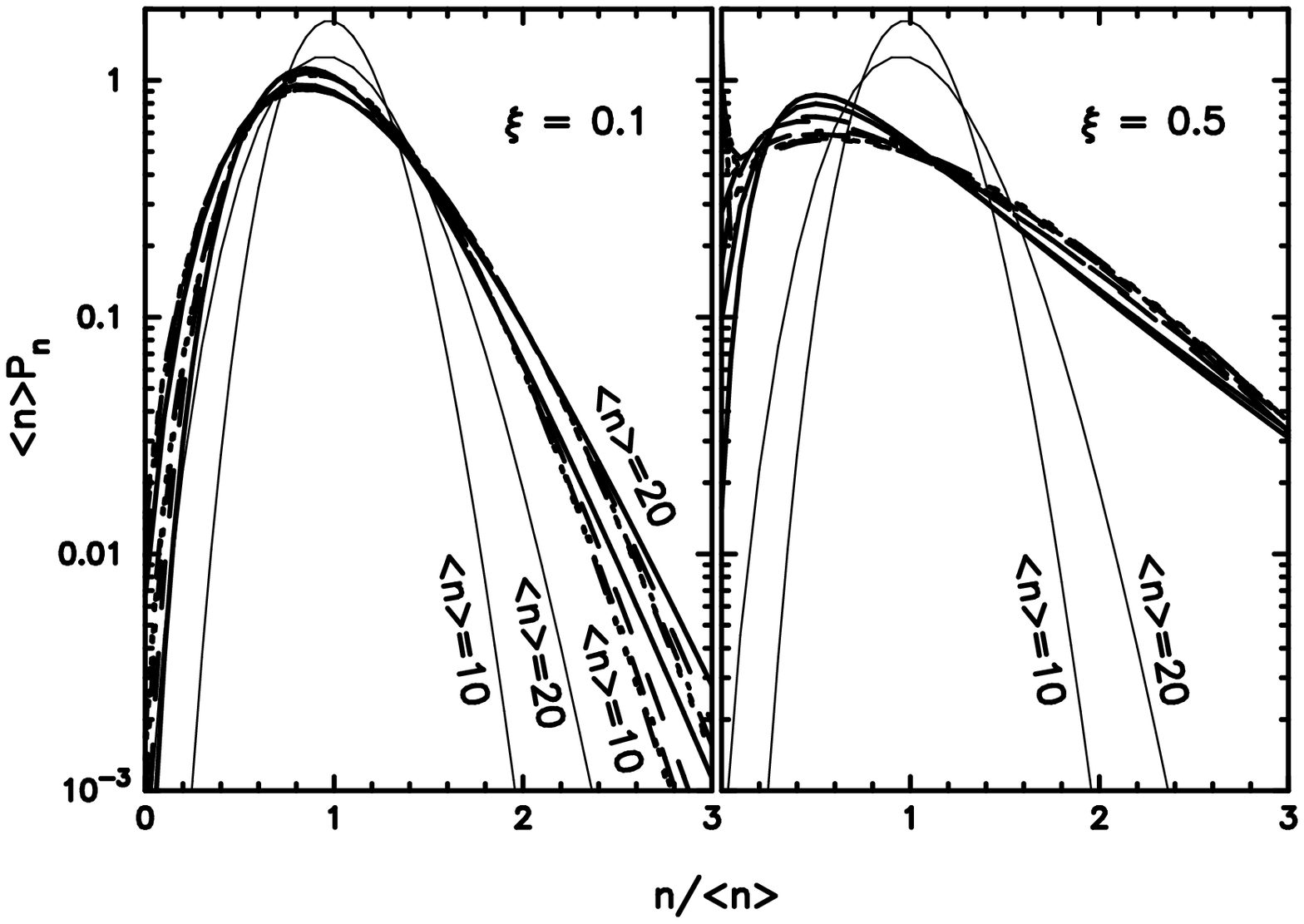}
}
\caption{KNO plot of $\bra n\ket P_n$ for fixed $\bra n\ket = 10$ and 20 
for $a = 1/2$, 1, 2, 3, 4. 
 }
 \label{knoscl}
\end{figure}

Once we know the generating function $Z(\vec x)$ or $f_0(\vec x)$
we may study the probability distribution  $P_n$ 
using Eq.(\ref{pax}) or the recurrence relation Eq.(\ref{recur});
\begin{eqnarray}
 P_n(x, z) &=& \frac{Z_n(x, z)}{Z(x, z)}
         = \frac{1}{Z(x,z)} \frac{z^n}{n!}
            \left[\left(\frac{d}{d z}\right)^n Z(x,z)\right]_{z=0}
\end{eqnarray}
for $n = \sum_k k n_k$.
For NB which is a special case of HGa in the $a \to 1$ limit,
 $x_k = x \frac{z^k}{k}$,
\begin{eqnarray}
 Z^{\rm NB}(x,z) &=& (1 - z)^{-x} = P_0^{-1}(x, z)    \\   
 P_n^{\rm NB}(x,z) &=& \frac{1}{Z} \frac{z^n}{n!} \frac{\Gamma(x+n)}{\Gamma(x)}
    = (1-z)^x \frac{z^n}{n!} \frac{\Gamma(x+n)}{\Gamma(x)}
             \label{nbprob}
\end{eqnarray}
The $P_n$ for various cases of Table \ref{tabl1} are shown
in Fig.\ref{cankno} and Fig.\ref{canpna} with the same mean value 
 $\bra n\ket = \bar n$ and the same fluctuation $\xi = \kappa_2 = f_2/\bar n^2$
for fixed $\bra n\ket = 10$.
Fig.\ref{knoscl} shows KNO plots of $\bra n\ket P_n$ versus $n/\bra n\ket$ 
for fixed $\bra n\ket = 10$ and 20.
Fig.\ref{cankno} shows that the various models considered here
have almost the same distribution for small fluctuation ($\xi = 0.01$)
and in this case they are very similar to a Poisson's distribution.
For $\xi = 0.05$ the models are similar to each other except 
for larger $n/\bar n$
even though they are different from a Poisson distribution.
For larger fluctuations such as $\xi = 0.5$, the models have very different
forms even if they have the same mean value and fluctuation.
Fig.\ref{canpna} shows that the probability distribution of these models 
differ in their form for fluctuations larger than $\xi \approx 0.2$.
Ref.\cite{poistrn} shows the total charge distribution in hadronic collisions.
They fit the data in the range of $\bar n = 6 \sim 13$ using a SN distribution.
The SN is a mixture of NB ($a=1$) and geometric ($a=2$) which are the
cases we have shown in the figures. To determine which model fits the
data best, we need to know the exact value of $\xi$ and even higher moments of
the data beside the mean value $\bar n$.

\begin{figure}[htb]
\centerline{
   \epsfxsize=5.5in   \epsfbox{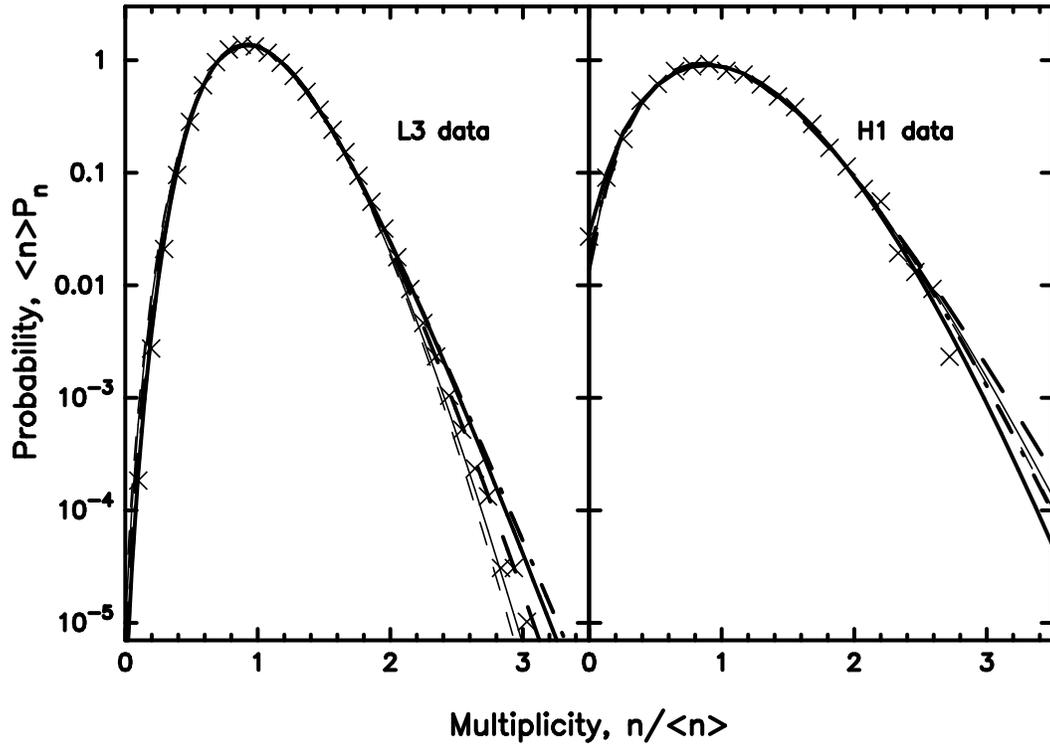}
}
\caption{Charged particle multiplicity distribution in jets of L3 data
\protect\cite{jeteel3} and H1 data \protect\cite{jeteph1}. 
The crosses are the data and the curves are the fits with HGa 
and GRW1D given in Table \protect\ref{tabldat}.
The thick solid curves are the GRW1D fit, the thick dashed curves are LC model,
the thin solid curves for NB model, the thin dashed curves for Geo model.
The dash-dotted curves are for HGa model with $a=0.31559$ for L3 data and
with $a=3.0$ for H1 data. 
 }
 \label{jetdat}
\end{figure}

Fig.\ref{jetdat} shows the multiplicity distribution for L3 data 
and H1 data discussed in Table \ref{tabldat}.
This shows that all the models having the same mean and fluctuation  
also have very similar fits to the data for $n/\bra n\ket$ smaller than 2.
For $n/\bra n\ket$ larger than 2, the NB (thin solid curve) or LC (thick dashed curve)
fits best the L3 data.
But for $n/\bra n\ket$ smaller than 1, the GRW1D (thick solid curve) and HGa 
with $a=0.31559$ (dash-dotted curve) fit the L3 data best and only these
two cases give the correct value of $\kappa_3$ for L3.
We can also see that the GRW1D (thick solid curves) fits best the H1 data 
for whole range of $n/\bra n\ket$. Only this GRW1D model fits the value 
of $\kappa_3$ for H1 data correctly. 
Thus we need to evaluate higher order moments of data (or at least up
to third order, $\kappa_3$) beside the
mean and fluctuation to understand the distribution of data 
and its underlying mechanism.
One difference between L3 and H1 data is that the H1 data has much larger probability 
for small $n/\bra n\ket$ compared to the L3 data. This large probability
at small $n/\bra n\ket$ causes the negative value for the third cumulant $\kappa_3$
for the H1 data.

The KNO behavior for $\bar n \to \infty$ can also be studied 
for the models listed in Table \ref{tabl1}.
According to KNO scaling, for distributions with a large mean value,
the distribution $\bra n\ket P_n$ becomes model independent in the new 
variable $n/\bar n$, i.e., variable scaled by mean value.
In general, any distribution becomes Gaussian for large mean $\bra n\ket$
according to the central limit theorem; specifically
\begin{eqnarray}
 P_n(\bar n, \sigma) &=& \frac{1}{\sqrt{2\pi\sigma^2}}
               \exp\left[-\frac{(n - \bar n)^2}{2\sigma^2}\right]   \nonumber  \\
   &=& \left(1 + \frac{1}{\xi \bar n}\right)^{-1/2}
          \frac{1}{\bar n \sqrt{2\pi\xi}}
               \exp\left[-\frac{1}{2\xi}\left(\frac{n}{\bar n} - 1\right)^2
              \left(1 + \frac{1}{\xi \bar n}\right)^{-1} \right] 
       \label{gauskno}
\end{eqnarray}
with the mean $\bra n\ket = \bar n$ and the standard deviation $\sigma$
which is related to the reduced factorial cumulant $\xi$ by
 $\sigma^2 = \bra n^2\ket - \bra n\ket^2 = \bar n + \xi {\bar n}^2$.
This means that the KNO scaling follows when the fluctuation is
given by $\sigma^2 = \xi {\bar n}^2$ with a constant $\xi$
or when $\xi \bar n \gg 1$.
Thus to compare KNO scaling properties of $\bar n P_n$ for different models,
the fluctuation of these models should have
the same value of $\xi = f_2/\bar n^2$.
For a small $\xi$, the Poisson component of the fluctuation
 $\sigma^2 = \bar n + \xi {\bar n}^2$
becomes dominant and thus the KNO scaling behavior is broken.
For large $\xi$ the Poisson component is negligible and 
KNO scaling is realized, i.e., 
\begin{eqnarray}
 \bar n P_n(\bar n, \xi) &=& \frac{1}{\sqrt{2\pi\xi}}
        \exp\left[-\frac{1}{2\xi}\left(\frac{n}{\bar n} - 1\right)^2\right]
     \label{knolim}
\end{eqnarray}
The KNO plot of Fig.\ref{knoscl} shows that the various distributions 
have no KNO scaling property for small fluctuation ($\xi = 0.1$)
but show a KNO scaling property for large fluctuation ($\xi = 0.5$).
The effect of mean on $\bar n P_n$ is larger than the difference 
between different model for $\xi = 0.1$ 
while the effect of mean on $\bar n P_n$ is much smaller than the difference 
between different models exhibiting KNO scaling behavior for $\xi = 0.5$.
Ref.\cite{poistrn} shows the total charge distribution in hadronic collisions
studied using SN model.
From their fits we can extract the corresponding fluctuation which
are $\xi = 0.05 \sim 0.5$ and $\bar n = 6 \sim 13$.
This means that the KNO scaling behavior is marginal for these data, i.e.,
just fitting the distribution $\bar n P_n$ of the data does not show clear 
evidence for KNO scaling.
We need to evaluate the explicit values of the mean number $\bar n$ and
the fluctuation $\xi$ to check the KNO behavior of this data;
the value of $\xi \bar n$ should be large enough to show KNO scaling
behavior as can be seen from Eqs.(\ref{gauskno}) and (\ref{knolim}).
Table \ref{tabldat} shows that $\xi < 0.1$ and $\xi \bra n\ket ~< 1$ for both data
indicating that the KNO scaling behavior would not be clearly exhibited
in jets for L3 and H1.

\subsection{Sequential procedures and compound Poisson distributions}
    \label{secseq}

A Poisson distribution plays a very important role in physics.
As already noted, in statistical physics, Maxwell-Boltzmann statistics 
leads to Poisson probabilities.
Other distributions are compared to the Poisson distribution
which acts as a benchmark for comparison.
The distributions considered in this paper can have large 
non-Poissonian fluctuations.
The purpose of this section is to show how they can be rewritten as
a compound process or sequential process involving one
aspect that has a Poisson character.
As an example the negative binomial distribution can be obtained
from a compound Poisson-logarithmic distribution as discussed
in Ref.\cite{poistrn}.
Here, we extended this result to include the other distributions
and we also show that the final distribution can be obtained
from compounding it with another distribution,
such as the negative binomial.
In general, the underlying picture for a sequential process
involves a two step procedure in which the observed particles
arise from the production of ``clusters'' with the
subsequent decay of each cluster producing its distribution
of particles.
The final distribution is obtained by compounding the 
probability distribution of the clusters with another distribution
coming from each cluster and suming over clusters.
Specifically,
the observed particles or members in system arise from production 
of $M = c$ clusters with probability distribution $P_c$.
This is sequentially followed by
each cluster decaying into $k_\alpha$ particles
with the probability $P_{k_\alpha}$
with $\alpha = 1$, 2, $\cdots$, $c$.
The probability of observing $n = \sum_k k n_k = \sum_\alpha k_\alpha$ 
particles is then obtained by a compound probability expression
\begin{eqnarray}
 P_n = \sum_c \sum_{\{k_\alpha\}} P_c \prod_{\alpha=1}^c P_{k_\alpha}
\end{eqnarray}
A negative binomial distribution can be obtained when
 $P_c = \bra c\ket^c e^{-\bra c\ket} / c!$
and $P_{k_\alpha} = (q^{k_\alpha} / k_\alpha) / \ln(1/p)$.
Here $c = M = \sum_k n_k$ and $\bra c\ket = \bra M\ket$,
 $p = (1-z) = (1+\bra n\ket/x)^{-1}$
and $q = 1-p = z = (\bra n\ket/x) / (1+\bra n\ket/x)$.
Also $N_c = \bra c\ket = x \ln(1 + \bra n\ket/x)$ and
 $n_c = \bra n\ket/N_c = (\bra n\ket/x) / \ln(1 + \bra n\ket/x)$.
This structure can be generalized as follow.

Since the generator of Poisson distribution is an exponential, i.e.,
the expansion of exponetial gives the Poisson's distribution
 $P_n^{\rm P}(\bar n)$, \begin{eqnarray}
 e^{\cal N} &=& \sum_{M=0}^\infty \frac{{\cal N}^M}{M!}
      = e^{\cal N} \sum_M e^{-{\cal N}} \frac{{\cal N}^M}{M!}
      = e^{\cal N} \sum_{M=0}^\infty P_M^{\rm P}({\cal N})
\end{eqnarray}
The grand partition function or the generating function $Z = e^{f_0}$ 
for any distribution can be represented as a Poisson distribution whose
mean value is the void variable $f_0$ or the grand potential $\Omega = - f_0$.
On the other hand and in general we can rewrite the void variable as
\begin{eqnarray}
 f_0(\vec x) &=& \ln Z(\vec x) = \sum_{k} x_k
     = {\cal N} \sum_{k} {\cal P}_k(\vec x)
\end{eqnarray}
where ${\cal N} = f_0 = \sum_k x_k$ and
\begin{eqnarray}
 {\cal P}_k(\vec x) &=& \frac{x_k}{\cal N}    
\end{eqnarray}
The ${\cal P}_k(\vec x)$ can be connected to its generating 
function ${\cal G}(\vec x, u)$:
\begin{eqnarray}
 {{\cal G}(\vec x, u)} &=& \sum_k (1-u)^k {\cal P}_k(\vec x)
    = \frac{1}{\cal N} \sum_k x_k (1-u)^k
\end{eqnarray}
Thus the generating function $G(\vec x, u)$ of $P_n$ can be expanded
in terms of ${\cal P}_k$ as
\begin{eqnarray}
 G(\vec x, u) &=& \sum_n (1 - u)^n P_n(\vec x)
      = \frac{1}{Z(\vec x)} \sum_n Z_n(\vec x) (1 - u)^n
     = \frac{e^{{\cal N G}(\vec x, u)}}{e^{\cal N}}
                      \nonumber \\
    &=& \sum_{M=0}^\infty \frac{1}{e^{\cal N}} \frac{[\cal N G]^M}{M!}
     = \sum_M P_M^{\rm P}({\cal N}) \left[\sum_j (1-u)^j {\cal P}_j\right]^M
                      \nonumber \\
     &=& \sum_M P_M^{\rm P}({\cal N})
               \sum_{\vec n_M} \prod_k [(1-u)^k {\cal P}_k]^{n_k}
\end{eqnarray}
where $P_M^{\rm P}({\cal N}) = e^{-{\cal N}} {\cal N}^M/M!$ is
the Poisson distribution with the mean of ${\cal N} = f_0$.
Here the sum over $\vec n_M$ is the sum over partitions $\vec n$
with a fixed $M = \sum_k n_k$.
Thus we have
\begin{eqnarray}
 P_n(\vec x) &=& \sum_M P_M^{\rm P}({\cal N})
          \sum_{\vec n_M} \prod_k {\cal P}_k^{n_k}(\vec x)   \label{comprb}
\end{eqnarray}
with $M = \sum_k n_k$ and $n = \sum_k k n_k$.
Any distribution obtained from a generating function of the form
of $Z(\vec x) = e^{f_0} = \exp[{\sum_k x_k}]$
can therefore be decomposed as a compound Poisson's distribution
with some other distribution ${\cal P}_k = x_k/f_0$ obtained from
the weight $x_k$. 

The sequential nature of a process is explicitly shown on Eq.(\ref{comprb}).
The observed particle multiplicity distribution arises from a
two step process in which $M = \sum_k n_k$ clusters are first 
distributed according to a Poisson distribution.
This is then sequentially followed by breaking each
of the $n_k$ clusters of type $k$ into $k$ particles
with probability ${\cal P}_k = x_k/{\cal N}$ and with $n = \sum_k k n_k$.
The probability associated with a given $M$ and $\vec n$ with $\vec x$
is $P_M(\vec x, \vec n) = P_M^{\rm P}({\cal N}) \prod_k {\cal P}_k^{n_k}
    = P_M^{\rm P}({\cal N}) \prod_k (x_k/{\cal N})^{n_k}$.

As an illustration we consider the LC model 
with $x_k = x C_k z^k/2^{2(k-1)}$.
Using the evolutionary variables \cite{prl86} $x = \beta/4p$ and $z = 4p(1-p)$
then ${\cal N} = \sum x_k = \beta$ for $p \le 1/2$ as already noted in Sect. \ref{ancest}
so that $P_M^{\rm P} = e^{-\beta} \beta^M/M!$.
The $x_k = \beta C_k p^{k-1} (1-p)^k$ so that
 ${\cal P}_k = x_k/{\cal N} = C_k p^{k-1} (1-p)^k$.
The underlying diagram associated with ${\cal P}_k$ are shown 
in Fig.\ref{LCfig}.
For a negative binomial (NB) distribution, $x_k = x z^k/k$ and 
thus ${\cal P}_k = x_k/{\cal N}$ is generated
from ${\cal N} = \sum_k x_k = - x \ln (1-z)$.
Therefore the NB is a compound Poisson-Logarithmic
distribution as shown in Table \ref{tabl3}.

As another example, 
we consider the HGa model with general $a$ instead of $a = 1/2$ for LC 
or $a = 1$ for NB.
The weight $x_k$ has the structure of the probability $P_k(x,z)$ of NB 
distribution given by Eq.(\ref{nbprob}), i.e.,
\begin{eqnarray}
 x_k &=& x \frac{z^k}{k!} \frac{\Gamma(a+k-1)}{\Gamma(a)}
    = \frac{x}{a-1} \frac{1}{(1-z)^{a-1}} P_k^{\rm NB}(a-1,z)     \label{xkpknb}
\end{eqnarray}
thus ${\cal P}_k = [1 - (1-z)^{a-1}]^{-1} P_k^{\rm NB}$ for HGa.
Therefore the HGa  $P_n$ distribution is a compound Poisson-NB distribution.
This may interpreted as a sequential process in which clusters with
a Poisson cluster distribution $P_c$ 
breakup into particles with a particle distribution ${\cal P}_k$
given by a NB distribution with parameter $a$.
For the various models considered here with their $x_k$ listed 
in Table \ref{tabl1},
the corresponding distribution ${\cal P}_k$ and the normalization 
factor ${\cal N} = f_0$ are listed in Table \ref{tabl3}.
We can further see that HGa can be looked as a 
compound Poisson-Poisson-Logarithmic distribution, i.e., 
a distribution having three sequential steps;
Poissonian breakup into clusters $\to$ Poissonian breakup of each 
cluster $\to$ logarithmic breakup of each of them.

\begin{table}[htb]
\caption{  
Poissonian sequential distribution for various models 
of Table \protect \ref{tabl1}.
 }    \label{tabl3}
\begin{tabular}{c|c|c|l} 
 \hline 
  Model  &  Weight ${\cal N}$ & Distribution $\ {\cal P}_k$
       &  Comments on ${\cal P}_k$  \\
 \hline 
  P    & $\bra n\ket$ & $\frac{1}{N}$ 
       & Monomer only or Uniform for $N$ species  \\  
  Geo  &  $x \frac{z}{1-z}$  & $(1-z) z^{k-1}$ 
       &  Uniform with constituents   \\
  NB     & $x \ln\left(\frac{1}{1-z}\right)$    &
           $\frac{z^k}{k} / \ln\left(\frac{1}{1-z}\right)$
       &  logarithmic with constituents  \\
  LC   & $2x \left[1 - \sqrt{1-z}\right]$  &
      $\left[\frac{1/2}{1 - \sqrt{1-z}}\right]
               \frac{z^k}{k!} \frac{\Gamma(k-1/2)}{\Gamma(1/2)}$
       &  NB with constituents with ${\cal P}_0 = 0$   \\
  HGa  & $\frac{x}{1-a} \left[1 - (1-z)^{1-a}\right]$  &
     $\left[\frac{(1-z)^{1-a}}{(1-z)^{1-a} - 1}\right] P_k^{\rm NB}(a-1, z)$ 
       &  NB with constituents without $k = 0$   \\
 GRW1D & $\frac{xz}{(1-z)^a}$   & $P_{k-1}^{\rm NB}(a, z)$
       &  NB with $x=a$    \\
 $ x_k = \frac{x}{k!} z^k$  &  $x e^{a z}$  &  $e^{-a z} \frac{z^k}{k!}$
       &  Poisson (exponential)    \\
 \hline     
\end{tabular}    
\end{table}

Because of a unique role played by the Poisson distribution
and the form $e^{f_0} = \exp[{\sum_k x_k}]$ of the generating function,
the cluster distribution $P_c$ is usually taken to be a Poisson.
However, as noted before, other divisions are possible.
Using the same approach used above for $P_c = P_M^{\rm P}$,
we can expand any distribution
using a NB instead of Poisson, i.e., 
with $P_c = P_M^{\rm NB}$  using
the form of the generating function $Z(x,z) = (1-z)^{-x}$ for the NB.
Replacing $z$ by ${\cal N}(z)$, 
the normalization factor of a new distribution, we have
\begin{eqnarray}
 Z(x,z) &=& \left[1 - {\cal N}(z)\right]^{-x}
     = \left[1 - {\cal N}(z) {\cal G}(u=0)\right]^{-x}
\end{eqnarray}
The ${\cal G}(u)$ is
\begin{eqnarray}
 {\cal G}(u) &=& \frac{{\cal N}([1-u]z)}{{\cal N}(z)}
    = \sum_k (1-u)^k {\cal P}_k
\end{eqnarray}
while the $G(u)$ is
\begin{eqnarray}
 G(u) &=& \sum_n (1 - u)^n P_n
    = \frac{Z(x, [1-u]z)}{Z(x, z)}
    = \frac{\left[1 - {\cal N} {\cal G}(u)\right]^{-x}}
           {\left[1 - {\cal N}\right]^{-x}}
\end{eqnarray}
Thus we have
\begin{eqnarray}
 G(u) &=& \sum_{M=0}^\infty (1 - {\cal N})^x \frac{{\cal N}^M}{M!}
         \frac{\Gamma(x+M)}{\Gamma(x)} [{\cal G}(u)]^M
    = \sum_{M=0}^\infty P_M^{\rm NB}(x, {\cal N}) [{\cal G}(u)]^M
         \nonumber  \\
  &=& \sum_{M=0}^\infty P_M^{\rm NB} 
             \left[\sum_{j=0}^\infty (1-u)^j {\cal P}_j\right]^M 
   = \sum_{M=0}^\infty P_M^{\rm NB}
        \sum_{\{n_k\}_M}  \prod_{k} \left[(1-u)^k{\cal P}_k\right]^{n_k}     \\
 P_n(\vec x) &=& \sum_M P_M^{\rm NB}(x, {\cal N}) 
        \sum_{\{n_k\}_M} \prod_k {\cal P}_k^{n_k}(\vec x)   \label{compnb}
\end{eqnarray}
The result of Eq.(\ref{compnb}) shows that
the distribution $P_n$ can be written as a compound probability distribution
of a negative binomial $P_M^{\rm NB}$ with another probability ${\cal P}_k$
distribution generated from ${\cal G}(u)$.
For the case of ${\cal N}(z) = e^z$, 
which may be considered as the fugacity $e^z = e^\mu$ for a particle
with chemical potential $\mu = z$,
${\cal G}$ can further be decomposed as
\begin{eqnarray}
 {\cal G}(u) &=& \frac{{\cal N}((1-u)z)}{{\cal N}(z)} = \frac{e^{(1-u)z}}{e^z}
    = \sum_{k=0}^\infty (1-u)^k P_k^{\rm P}(z)
\end{eqnarray}
i.e., ${\cal P}_k$ for this case is Poisson $P_k^{\rm P}(z)$.
If ${\cal N} = 1 - e^{f_0/x}$, then ${\cal P}_k = P_k^{\rm P}(f_0/x)$
without $k = 0$ and $Z(x,z) = \left[1 - (1-e^{f_0/x})\right]^x = e^{f_0}$.
For $f_0 = \sum_k x_k$ given in Table \ref{tabl1},
the ${\cal P}_k$ becomes the same probability with $x_k$ replaced
by $x_k/x$. As an example the HGa with
 $Z(x,z) = \exp\left(\frac{x}{1-a}[1 -(1-z)^{1-a}]\right)$
can be decomposed as a sequential process consisting of a NB 
distribution of clusters with $Z(x,{\cal N}=1-e^{f_0/x})$ 
which is then followed by a breakup of clusters distributed with
a HGa distribution 
with $Z(x=1,z)$ but without voids, i.e., with ${\cal P}_0 = 0$.
Thus this decomposition separates the parameter $x$ assigned to cluster 
from other parameters.

\section{conclusion}

Event-by-event studies from 
high energy collisions
are being used to study the details of
particle multiplicity distributions as, for example,
those associated with pions.
Such studies not only give information about the mean number of
particles produced, but also information about fluctuations
and higher order moments of the probability distribution
which are important tools for studying the underlying processes
and mechanisms that operate.
They are also useful in distinguishing various phenomenological models.
Issues associated with fluctuations play an important role in
many areas of physics and departures from Poisson statistics are 
of current interest.
One purpose of this paper was an investigation of various models of
particle multiplicity distributions that can be used in
event-by-event analysis.
These various phenomenological models are developed using a general form
of a unified model which is based on a grand canonical partition function
and an underlying weight arising from Feynman's path integral approach
to statistical processes. A resulting distribution has three control
parameters called $a$, $x$, $z$. The relationships of these parameters to various
physical quantities are discussed. One important result is the
connection of the parameter $a$ to the Fisher exponent $\tau$; namely $\tau = 2-a$
for HGa.
This connection arises from a parallel we developed between the model for
particle multiplicity distributions considered here and our previous
approach to cluster yields. Since an exact description of particle
multiplicity distributions is not known, we have considered several cases
with different $\tau$'s or $a$'s which are contained in our unified
description. Moreover, many of the existing distributions currently used
in particle phenomenology are shown to be special choices of $\tau$ or $a$
which appear in a generalized hypergeometric model called HGa. These
include the Poisson distribution coming from coherent emission, chaotic
emission producing a negative binomial distribution, combinations of
coherent and chaotic processes leading to signal/noise distributions and
field emission from Lorentzian line shapes producing the Lorentz/Catalan
distribution. Using the HGa model combinants, cumulants and moments are
discussed and a physical significance is given to combinants in terms of
the underlying partition weights of a Feynman path integral approach to
statistical processes. The parameter $a$ or Fisher exponent $\tau$ is shown
to play an important role in the behavior of the combinants which
manifest itself in various physical relationhips. The HGa model and its
associated special cases are used to explore a wide variety of phenomena.
These include: linked pair approximations leading to hierarchical scaling
relations on the reduced cumulant level, generalized void scaling
relations, clan variable descriptions and their connections with
stochastic variables and branching processes, KNO scaling behavior,
enhanced non-Poissonian fluctuations. Models based on an underlying random
walk description are also discussed.

  In this paper we compared various particle multiplicity distributions
within the hypergeometric model HGa. Our results show that even though
various distributions have the same mean and fluctuation, the distribution
itself or the underlying mechanism could be different. Comparisons within
the HGa model also show that just comparing void variables $\chi$ and $\xi$
or mean $\bar n$ and fluctuation $f_2$ or $\sigma$
is not enough to distinguish different models that descibe particle
multiplicity data. Thus, to find the correct distribution and underlying
mechanism from various data more information than just the mean and
fluctuation are necessary and new variables should be found which are
quite different between different models. For example, higher order
reduced factorial cumulants need also to be evaluated such as the third
order cumulant $\kappa_3$.

  Applications of our approach to the charged multiplicity data of L3
and H1 are given. The mean $\langle n\rangle$, fluctuation variable $\xi$, void scaling
variable $\chi$, and third order reduced cumulant variable $\kappa_3$  obtained
from these experimnets are compared with various models discussed in this
paper. The value of the third cumulant $\kappa_3$ shows that the GRW1D model
can fit the charge multiplicity data of L3 and H1. 
The HGa model also gives a fit to the L3
data, but is not as successful as GRW1D with respect to the
third order reduced cumulant of the H1 data. Both have very
similar multiplicity distributions as shown in Fig.\ref{jetdat}
which compares the two models with the charged particle multiplicity
distributions of jets for L3 and H1 data. Differences appear
in the high multiplicity events with reduced multiplicity
$n/\langle n\rangle > 2.5$.
The reduced third cumulant is negative for the H1 data of
Table \ref{tabldat}.
Negative values of a cumulant require negative values of the
parameter $a$ which are allowed in GRW1D if the parameters $x$
and $z$ are also taken as negative, but a negative value of $a$ is
not acceptable in HGa since this would require a value of $z$
greater than 1 and thus a complex $x$.
We also discuss the KNO
scaling behavior for these data and conclude that the KNO scaling behavior
would not be clearly exhibited in the jets for L3 and H1.

In this paper we have also generalized 
the compound distribution that arises from sequential process 
which may reveal the dynamical structure of the distribution.
Specifically, the underlying sequential picture involves a two step process
where the final distribution arises from the production of clusters
followed by a subsequent decay of the clusters.
For the HGa model, the final distribution is obtained from 
compounding a Poisson distribution of clusters with a NB distribution
coming from the decay of each of the clusters.
The HGa may arise through a three step sequential process of
Poisson-Poisson-Logarithmic compound distribution.
It is also shown that the HGa can arise from
a two step sequential process of a NB distribution followed by a new HGa 
with a different mean value.

This work was supported in part by Grant No. 2001-1-11100-005-3 
from the Basic Research Program of the Korea Science and
Engineering Foundation
and in part by the DOE Grant No. DE-FG02-96ER-40987.

\end{document}